\documentclass[prd,aps,tightenlines,floats,nofootinbib,preprintnumbers]{revtex4-2}
\usepackage{graphicx}
\usepackage{amsmath,amssymb}
\usepackage{grffile}
\usepackage{ulem}
\usepackage{soul}
\usepackage[colorlinks=true,allcolors=blue]{hyperref}
\usepackage[capitalise]{cleveref}
\usepackage{multirow}

\usepackage[dvipsnames]{xcolor}

\newcommand{\df}[2]{\frac{d#1}{d#2}}
\newcommand{\ddf}[2]{\frac{d^2#1}{d#2^2}}

\newcommand{\dd}{\partial}

\newcommand{\xx}{\boldsymbol{x}}

\newcommand{\rr}{\boldsymbol{r}}

\begin{document}

\preprint{RUP-23-28}
\title{Dynamical simulations of colliding superconducting strings}

\author{Takashi Hiramatsu}\email{hiramatz@rikkyo.ac.jp}
\affiliation{Department of Physics, Rikkyo University\\ Toshima, Tokyo 171-8501, Japan
}

\author{Marc Lilley}\email{marc.lilley@obspm.fr}
\affiliation{SYRTE, Observatoire de Paris, Universit\'e PSL, CNRS,
Sorbonne Universit\'e, LNE, 61 avenue de l’Observatoire 75014 Paris, France}

\author{Daisuke Yamauchi}\email{d-yamauchi@ous.ac.jp}
\affiliation{
Department of Physics, Faculty of Science, Okayama University of Science\\ 1-1 Ridaicho, Okayama, 700-0005, Japan
}

\begin{abstract}
We study the collisions of elastic superconducting strings, also referred to as current-carrying strings, formed in a $U_{\rm local}(1)\times U_{\rm global}(1)$ field-theory model, using three-dimensional numerical field-theoretic simulations. The breaking of $U_{\rm local}(1)$ leads to string formation via the Higgs mechanism, while the scalar field of the second $U_{\rm global}(1)$ carries the current, which condenses onto the string.  We construct straight and static superconducting string solutions numerically and identify the regions in which they exist in the model parameter space. We then perform dynamical simulations for colliding superconducting strings with various collision angles and collision velocities. We explore the kinematic parameter space for six sets of model parameters characterising the coupling between the two scalar fields and the current on the string.
The final states of the strings (after the collision) are reported diagrammatically. We classify them into four categories: (i) {\it regular} intercommutation, (ii) {\it double} intercommutation, (iii) bound state, and (iv) expanding string solution. We find that the outcome of the collision process is the regular intercommutation of the colliding strings in most of the kinematic parameter space while they form bound states for small velocities and small angles. We also find that the strings undergo two successive intercommutations and, therefore, pass through one other in a small region corresponding to relatively small angles and velocities of order $c/2$. The string structure breaks down when there is a relatively large coupling between the two scalar fields, even if each string is stable before the occurrence of the collision.

\end{abstract}

\maketitle

\section{Introduction}
\label{sec:intro}

Cosmic strings can form in many models of grand-unification, see, e.g.,~\cite{Jeannerot:2003qv}, and hence, determining their observational consequences --- be it via gravitational waves, lensing, cosmic microwave background physics, or any other means --- is of great interest~\cite{VilenkinShellard,Kibble:1976sj, Hindmarsh:1994re, Copeland:2009ga}. Indeed, if any signature of cosmic strings was to be observed (so far, only bounds exist), it would provide invaluable information about the Universe during its phase transitions and open up a window on fundamental physics at very high energies, such as axion physics \cite{Davis:1986xc, Dabholkar:1989ju, Gorghetto:2018myk, Kawasaki:2014sqa, Hiramatsu:2010yn, Kawasaki:2018bzv, Yamaguchi:1998iv, Yamaguchi:1999dy, Yamaguchi:2002sh}, grand-unification \cite{Davis:1997bs, Jeannerot:2003qv, Cui:2007js, Allys:2015kge} and the seesaw mechanism \cite{Dror:2019syi, Samanta:2020cdk, Fu:2023nrn}.

Central to any calculation of the observational consequences of cosmic strings is that the string network reaches a ``scaling solution'', in which the energy density in the strings is a small fixed fraction of the energy density in the Universe.  Most of the work related to string networks has been based on the Abelian-Higgs model for cosmic strings; in this particular case, strings generally intercommute when colliding (see below). This crucial fact means that the string network will form closed loops during its evolution; these loops radiate energy in the form of gravitational waves (see, e.g., \cite{Sousa:2014gka, Blanco-Pillado:2017oxo, Ringeval:2017eww, Sousa:2016ggw, Auclair:2019wcv, Allen:1991bk, Durrer:1989zi, Garfinkle:1988yi, Burden:1985md, Vachaspati:1984gt, Garfinkle:1987yw, Vilenkin:1981bx, Binetruy:2012ze, Sanidas:2012ee, Kuroyanagi:2012wm, Blanco-Pillado:2013qja, Sousa:2020sxs, Blanco-Pillado:2017oxo, Lorenz:2010sm, Blanco-Pillado:2019tbi, Sousa:2014gka, LIGOScientific:2021nrg, Dufaux:2010cf, Sousa:2013aaa, Bian:2022tju, Kitajima:2022lre, Vachaspati:1984yi, Blanco-Pillado:2017rnf, CamargoNevesdaCunha:2022mvg, Chen:2022azo, Lozanov:2023aez, Auclair:2022ylu, Rybak:2022sbo, Siemens:2006yp, Jenkins:2018nty, Gelmini:2021yzu, Damour:2000wa, Damour:2001bk, Kitajima:2023vre, Damour:2004kw}), and it is for this reason that the network evolves towards an attractor scaling solution (see, e.g., \cite{Kibble:1984hp,Martins:1995tg,Martins:1996jp,Martins:2000cs,Hiramatsu:2013tga,Hindmarsh:2017qff,Hindmarsh:2018wkp,Correia:2018gew}).
In this regime, there is a unique length scale, called the correlation length. The (velocity-dependent) one-scale model was developed with this particular property in mind \cite{Kibble:1984hp, Bennett:1985qt, Albrecht:1989mk, Austin:1993rg, Martins:1996jp, Martins:2000cs, Vanchurin:2013tk}, and allows one to estimate the present power spectrum of the gravitational waves generated by cosmic strings as well as the contribution of the cosmic string network to the cosmic microwave background \cite{Lazanu:2014eya, Lazanu:2014xxa, Lizarraga:2016onn, Charnock:2016nzm, Rybak:2017yfu, Rybak:2021scp, Hindmarsh:2018wkp, Landriau:2002fx, Bevis:2006mj, Daverio:2015nva, Lizarraga:2016onn, Lopez-Eiguren:2017dmc, Ramberg:2022irf, Silva:2023diq}.

Generally, the outcome of a collision of two U(1) Abelian-Higgs cosmic strings depends on the relative velocity $v$ of the colliding strings, their relative angle $\alpha$, and of course, on whether the strings are in the Type-I or Type-II regime \cite{VilenkinShellard}.  In the Type-II regime, no bound states can form (in other words, a string with winding 2 has higher energy per unit length than two strings with winding 1). In that case, extensive studies have shown that the strings intercommute -- that is, they exchange partners -- in most of the $(\alpha,v)$ plane, except for the $v\rightarrow 1$ case, where double reconnections occur \cite{Achucarro:2006es,Achucarro:2010ub}.
In the Type-I regime or for cosmic superstrings, bound states with corresponding Y-junctions can form \cite{Rybak:2018oks,Avgoustidis:2014rqa,Bevis:2008hg,Copeland:2003bj,Jackson:2004zg,Polchinski:1988cn,Binetruy:2010bq,Matsui:2020hzi,Steer:2017xgh}, thus leading to a more diverse set of final states.  Analytic studies of these collisions have shown that at low $v$ and $\alpha$, bound states are {\it kinematically} allowed to form while numerical simulations have shown that they generally form {\it dynamically}, see \cite{Salmi:2007ah}.

In this paper, we extend previous studies to the collisions of $ U(1)_{\rm local}\times  U(1)_{\rm global}$ strings, in which the second $U(1)_{\rm global}$ leads to the formation of currents on the strings. 
The motivation for this work is precisely that in GUT models, such current-carrying cosmic strings, also referred to as ``superconducting strings'', are thought to form generically \cite{Witten:1984eb, Jeannerot:2003qv}. Superconducting strings and their physical properties have been studied extensively in the literature \cite{Davis:1988ij, Martins:1998gb, Carter:1996rn,Babul:1987me, Everett:1988tn, Davis:1996xs, Peter:1993tm, Garaud:2009uy, Davis:1995kk, Lilley:2010av, Fukuda:2020kym, Abe:2020ure, Kibble:1996rh, Peter:1992dw, Peter:1992ta, Hartmann:2016axn, Hartmann:2017lno,Lilley:2009yr, Oikonomou:2010ut, Babeanu:2011ie, Allen:1995rd, Copeland:1987th, Cordero-Cid:2002hmv, Blanco-Pillado:2002vwq, Dimopoulos:1997xa, Martins:1997nb, Martin:1998cv, Dimopoulos:1999dn, Larsen:1993ha, Martins:1999jg}.
As in the case of standard cosmic strings, semi-analytic models of superconducting strings have been developed~\cite{Martins:1999ja, Oliveira:2012nj, Martins:2014kda, Vieira:2016vht, Rybak:2017yfu, Rybak:2023jjn, Martins:2020jbq, Martins:2021cid, Rybak:2018oks, Rybak:2020pma}, and the radiation of gravitational waves~\cite{Rybak:2022sbo, Auclair:2022ylu, Babichev:2003au, Babichev:2002bn} and the radiation of electromagnetic waves~\cite{Blanco-Pillado:2000nbp, Miyamoto:2012ck, Imtiaz:2020igv, Garfinkle:1987yw, Copeland:1987yv, Blanco-Pillado:2000ssd, Spergel:1986uu,Ferrer:2005xva,Vilenkin:1986zz,Tashiro:2012nv} by superconducting string networks have been studied.   It is of interest to find out whether the network of such strings follows a scaling law, as Abelian-Higgs cosmic strings do. The key to obtaining such a scaling law is that the strings can efficiently reconnect with each other to create loops. Here, we focus on the collision of two current-carrying strings using field-theoretic numerical simulations. Recently, the authors in \cite{Fujikura:2023lil} have investigated the interaction between two straight current-carrying strings and revealed that an additional (gauged) scalar field becomes an additional source of attraction. Thus, the dynamics of such strings may differ from those of Abelian-Higgs strings.

Collisions of current-carrying strings will depend not only on ($\alpha,v$) but also on the current on each string.  In a previous paper~\cite{Steer:2017xgh}, we have studied these collisions analytically using the well-known elastic string model \cite{Witten:1984eb} applied to the case of electric and magnetic strings. In the elastic string model, for an infinite string in a stationary state, the energy per unit length of the string $U$ is related to its tension $T$ by a barotropic equation of state $U(T)$.  In~\cite{Steer:2017xgh}, we showed that this description breaks down when electric or magnetic current-carrying strings collide to form a bound state, which we interpret as being due to the inability of the elastic string model to take into account non-conservative processes, as, for instance, the appearance of dissipative processes at the Y-junction. In \cite{Rybak:2018oks} and \cite{Rybak:2020pma}, the authors performed a similar analysis for strings carrying chiral currents and for transonic strings. In both cases, by contrast to the general case, a wave-like solution exists, and the authors were able to show that a bound state with Y-junctions exists in the context of the elastic string model.

Hence, unlike in the Nambu-Goto case (without currents), it would appear to be impossible -- at least with simple models -- to determine the outcome of a collision between either electric or magnetic current-carrying strings.  For this reason, we present here, as an alternative, a numerical study of the same situation: the collision of two straight current-carrying elastic strings in a $U_{\rm local}(1)\times U_{\rm global}(1)$ field-theory model. We focus on the role of the current in the collision and study the outcome of collisions as a function of the couplings in the field theory and as a function of the collision velocity and angle.

This paper is organized as follows. In~\cref{sec:NOV}, we briefly introduce the field theory model used to describe current-carrying cosmic strings and outline how to determine the one-dimensional current-carrying vortex solution. We discuss the region of parameter space in which current-carrying strings can form and then numerically obtain straight and static superconducting string solutions. In~\cref{sec:setup}, we discuss the setup used to perform the field theory simulations for colliding superconducting strings. In~\cref{sec:results}, we show the numerical results of the colliding superconducting strings and summarise the final states resulting from the collision. In~\cref{sec:phase}, 
we classify the final states in the plane defined by the collision velocity and the collision angle.  Finally, we conclude in~\cref{sec:conclusion}.

\section{Superconducting string model and vortex solution}
\label{sec:NOV}

%%%%%%%%%%%%%%%%%%%%%%%%%%%
\subsection{Model}
\label{subsec:model}
%%%%%%%%%%%%%%%%%%%%%%%%%%%

We study a 
$U(1)_{\rm local}\times U(1)_{\rm global}$ 
current-carrying string model~\cite{Peter:1992dw} defined by the action
\begin{equation}
 S = -\int\!d^4x\,\sqrt{-g}\left(
   (D_\mu\varphi)^*(D^\mu\varphi)
 + (\partial_\mu\sigma)^*(\partial^\mu\sigma)
 + \frac{1}{4}F_{\mu\nu}F^{\mu\nu}
 + V(\varphi,\sigma)\right)\,,
 \label{eq:action}
\end{equation}
where $\varphi$ is a complex scalar field, and $F_{\mu\nu} \equiv \partial_\mu A_\nu - \partial_\nu A_\mu$ is the field strength, with $A_\mu$ the $U(1)$ gauge field. The covariant derivative is defined as $D_\mu \equiv \partial_\mu - ieA_\mu$.
The additional scalar field $\sigma$ which has a global $U(1)$ symmetry, will be referred to as the {\it current-carrier}. In Ref.~\cite{Laguna:1990it}, the authors consider a local $U(1)$ symmetry for the current-carrier, while in Ref.~\cite{Abe:2022rrh}, the model considered is similar to the one studied here, but using global strings with a gauged current-carrier instead.

We use the same potential $V(\varphi,\sigma)$ as in Ref.~\cite{Peter:1992dw}, namely
\begin{equation}
V(\varphi,\sigma) = 
  \frac{\lambda_\varphi}{4}(|\varphi|^2-\eta^2)^2
  + \lambda_{\varphi\sigma}(|\varphi|^2-\eta^2)|\sigma|^2
  + \frac{m_\sigma^2}{2}|\sigma|^2
  + \frac{\lambda_\sigma}{4}|\sigma|^4.
\label{eq:defV}
\end{equation}
The first term is the familiar Mexican-hat potential (with the vacuum expectation value, $\eta$), which gives rise to vortex solutions for $\varphi$. The second term couples the string forming field to the current-carrier $\sigma$, and thus plays an important role in the current condensation on the string. The correspondence with the notations in Ref.~\cite{Peter:1992dw} is summarised in~\cref{appsec:notations}. 

We consider the model in Minkowski space. In Lorenz gauge, the field equations read
\begin{align}
&\ddot{\varphi}
-\dd^i\dd_i\varphi
   +2ieA^\mu\dd_\mu\varphi
   +e^2A^\mu A_\mu\varphi
+ \frac{dV}{d\varphi^*} = 0
  \label{eq:eqphi} \\
&\ddot{A}_{\nu} - \partial^j\partial_j A_{\nu}
   = 2e{\rm Im}(\varphi^* \partial_\nu\varphi)-2e^2A_\nu|\varphi|^2,
  \label{eq:eqA}  \\
&\ddot{\sigma} - \partial^i\partial_i\sigma + \frac{dV}{d\sigma^*} = 0,
\label{eq:eqsigma}
\end{align}
with the dot denoting a derivative with respect to time. 

%%%%%%%%%%%%%%%%%%%%%%%%%%%%%%%%%%%%%%%%%%%%%%%%%%%%%%%
\subsection{Vortex solutions}
\label{subsec:1-vortex}
%%%%%%%%%%%%%%%%%%%%%%%%%%%%%%%%%%%%%%%%%%%%%%%%%%%%%%%

We use cylindrical coordinates and look for axially-symmetric Abrikosov-Nielsen-Olsen (ANO) vortex solutions\footnote{For the Abelian-Higgs model, see Refs.~\cite{Abrikosov:1956sx,Nielsen:1973cs}}. Following Ref.~\cite{Peter:1992dw}, we use the ansatz
\begin{align}
\varphi(\rr) &= \eta f(r)e^{in\theta}, \label{eq:no_phi} \\
A_i(\rr) &= -\epsilon_{ij}x_j\frac{n}{er^2}\alpha(r), \label{eq:no_a} \\
\sigma(t,\rr) &= \eta g(r)e^{i(kz-\omega t)}, \label{eq:no_sigma} 
\end{align}
where $f,\,g$ and $\alpha$ are dimensionless functions of the radial coordinate, $r$.
The first two expressions are exactly equivalent to those in the Abelian-Higgs model. Substitution into the field equations (\ref{eq:eqphi})-(\ref{eq:eqsigma}), we obtain
\begin{align}
 &\ddf{f}{x}
   + \frac{1}{x}\df{f}{x}
   - \frac{n^2f}{x^2}(\alpha-1)^2
   = \beta_\varphi(f^2-1)f
   + 2\beta_{\varphi\sigma}g^2f, \label{eq:f} \\
 &\ddf{\alpha}{x} - \frac{1}{x}\df{\alpha}{x}
    = 2f^2(\alpha-1), \label{eq:a} \\
 &\ddf{g}{x}
   + \frac{1}{x}\df{g}{x}
   - \gamma g
   = 2\beta_{\varphi\sigma}(f^2-1)g
     + \beta_\sigma g^3, \label{eq:g}
\end{align}
where 
\begin{equation}
 \beta_\varphi \equiv \frac{\lambda_\varphi}{2e^2}, \quad
 \beta_{\varphi\sigma} \equiv \frac{\lambda_{\varphi\sigma}}{2e^2}, \quad
 \beta_\sigma \equiv \frac{\lambda_\sigma}{2e^2}, \quad
 \mu^2_\sigma = \frac{m_\sigma^2}{2e^2\eta^2}, \quad
 K^2 = \frac{k^2}{e^2\eta^2}, \quad
 \Omega^2 = \frac{\omega^2}{e^2\eta^2}, \quad
 x = e\eta r.
\end{equation}
and 
$\gamma = K^2-\Omega^2+\mu^2_\sigma$.
In Ref.~\cite{Peter:1992dw}, the author introduced a parameter
$w\equiv k^2-\omega^2$ that becomes $w = e^2\eta^2(\gamma-\mu_\sigma^2)$
in our notation.
For later convenience, we define the conserved current of the current-carrier,
\begin{align}
 j^\sigma_\mu &= 2{\rm Im}\;(\sigma^*\partial_\mu\sigma).
\label{eq:def_j}
\end{align}

In the setup considered here, the current only has a $z$-component, $j_z^\sigma (r) = 2k\eta^2 g(r)^2 = 2eK\eta^3g(r)^2$. The parameter $K$ thus characterises the strength of the current.

%%%%%%%%%%%%%%%%%%%%%%%%%%%%%%%%%%%%%%%%%%%%%%%%%%%%%%%%%%%%%%%%%%%%%%%%%%%%%%%
\subsection{Conditions on model parameters for superconducting string formation}
\label{subsec:condition}
\begin{figure}[!t]
\includegraphics[width=8.5cm]{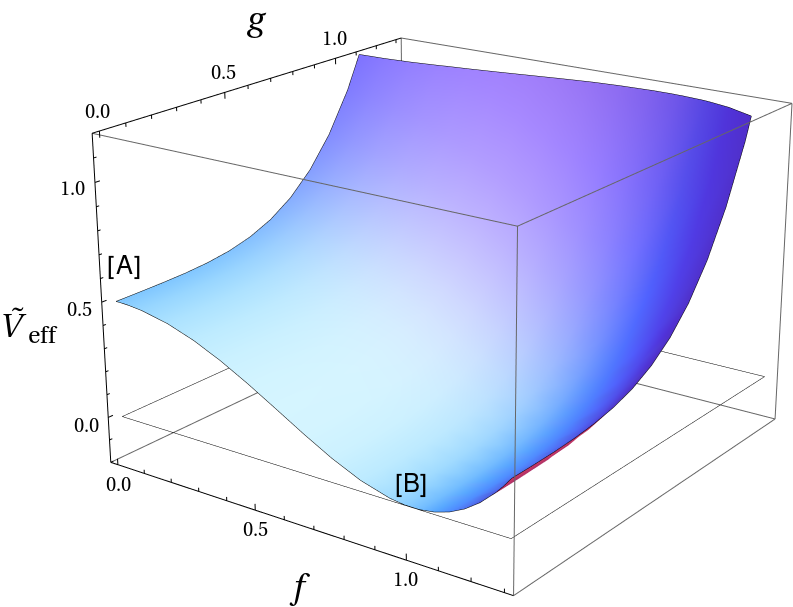}
\includegraphics[width=8.5cm]{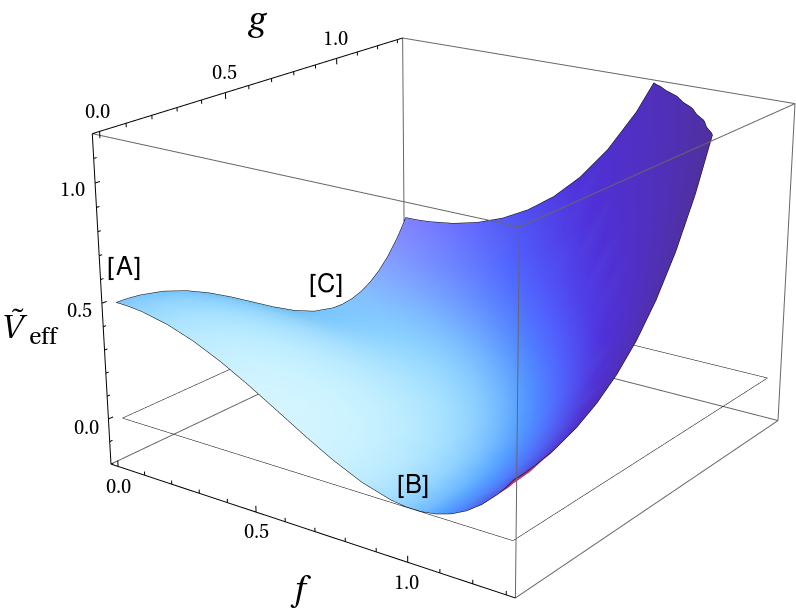}
\includegraphics[width=8.5cm]{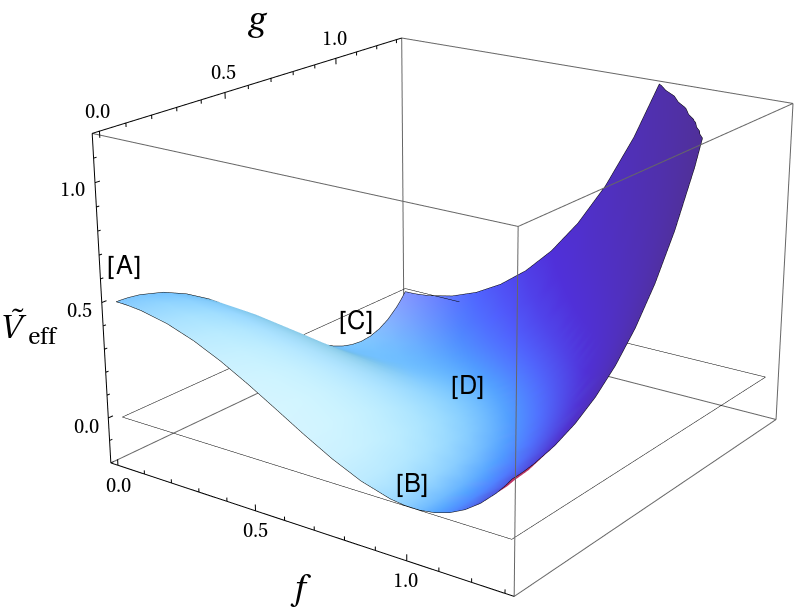}
\caption{Effective potential given in Eq.~(\ref{eq:Veff}) with
 $\beta_{\varphi\sigma}=0.02$ (upper-left plot), 0.49 (upper-right plot) and 0.60 (lower plot). We fixed $(\beta_\varphi,\beta_\sigma,\gamma)=(1.0,1.0,0.05)$. The labels [A], [B], [C] and [D] indicate the locations of the local extrema given in Eq.~(\ref{eq:extrema}).}
\label{fig:potential}
\end{figure}

%%%%%%%%%%%%%%%%%%%%%%%%%%%%%%%%%%%%%%%%%%%%%%%%%%%%%%%%%%%%%%%%%%%%%%%%%%%%%%%
We can obtain vortex solutions on which the current-carrier $\sigma$ condenses, i.e.~{\it superconducting strings}, only when the model parameters are in a certain region of parameter space.  Here we determine under which conditions the superconducting strings exist.  Substituting Eqs.~(\ref{eq:no_phi})-(\ref{eq:no_sigma}) into~\cref{eq:action}, we obtain
\begin{align}
-\frac{1}{e^2\eta^4}\mathcal{L}
 =
         \left(\frac{df}{dx}\right)^2
       + \frac{n^2}{x^2}f^2(1-\alpha)^2
       + \left(\frac{dg}{dx}\right)^2
       + \frac{n^2}{2x^2}\left(\frac{d\alpha}{dx}\right)^2
       + (K^2- \Omega^2)g^2
       + \frac{V}{e^2\eta^4}\,.
 \label{eq:red_Lag}
\end{align}
The effective potential reads
\begin{align}
  \widetilde{V}_{\rm eff} &\equiv \frac{V}{e^2\eta^4}+(K^2-\Omega^2)g^2, \notag\\
  &=  \frac{\beta_\varphi}{2}(f^2-1)^2
      + 2\beta_{\varphi\sigma}(f^2-1)g^2
      + \frac{\beta_\sigma}{2}g^4
      + \gamma g^2,\label{eq:Veff}
\end{align}
where again $\gamma \equiv K^2-\Omega^2+\mu_\sigma^2$. This effective potential determines whether a superconducting string can exist or not.  The shape of this potential is shown in \cref{fig:potential} for three distinct cases with $(\beta_\varphi,\beta_\sigma, \gamma)=(1.0,1.0,0.05)$ and $\beta_{\varphi\sigma}=0.02, 0.49$ and $0.60$.

The four critical points of the effective potential are given by
\begin{align}
 (f,g) = (0,0),\; (1,0),\; \left(0,\sqrt{\frac{2\beta_{\varphi\sigma}-\gamma}{\beta_\sigma}}\right),\;\left(
      \sqrt{1-\frac{2\beta_{\varphi\sigma}\gamma}{4\beta_{\varphi\sigma}^2-\beta_\sigma\beta_\varphi}},
        \sqrt{\frac{\beta_{\varphi}\gamma}{4\beta_{\varphi\sigma}^2-\beta_\sigma\beta_\varphi}}
     \right).
\label{eq:extrema}
\end{align}
We label them [A] to [D] in~\cref{fig:potential}. 
When $g=0$, there is no condensate, which occurs at locations [A] and [B]. 
The field profile $f(r)$ for a string carrying no current connects [A] and [B].
The field $g$ is non-zero at [C] and at the saddle point [D]. [C] corresponds to the desired vacuum state describing the string core on which the current-carrier $\sigma$ condenses, i.e. $\varphi=0$ and $|\sigma|>0$. [D] is a saddle point, i.e.~one eigenvalue of the Hessian matrix is positive while the other is negative. [C] and [D] do not exist depending on the choice of parameters.

To obtain a superconducting string configuration, we first impose $\widetilde{V}_{\rm eff}(A)>\widetilde{V}_{\rm eff}(B)$. This means that $\beta_{\varphi}>0$ is a necessary condition for spontaneous symmetry breaking. A stable superconducting string configuration also requires the existence of the extremum [C]. This condition yields $(2\beta_{\varphi\sigma}-\gamma)/\beta_\sigma>0$, see~\cref{eq:extrema}.  This condition is satisfied in the upper-right and lower plots of Fig.~\ref{fig:potential} while it is not satisfied in the upper-left plot.

In order for the condensate to form onto the string, the potential energy at [C],
\begin{align}
 \widetilde{V}_{\rm eff}(C) &=
  \frac{\beta_\varphi\beta_\sigma-(2\beta_{\varphi\sigma}-\gamma)^2}{2\beta_\sigma}.
 \label{eq:cond3}
\end{align}
should be smaller than that at [A], i.e.~$\widetilde{V}_{\rm eff}(A)=\beta_\varphi/2>\widetilde{V}_{\rm eff}(C)$. This means that $\beta_\sigma>0$. Using this condition and the one required to guarantee the existence of [C], we obtain $2\beta_{\varphi\sigma}>\gamma$. On the other hand, the current-carrier $\sigma$ should vanish far from the string core, requiring [B] to be a minimum. In other words, the Hessian matrix at [B] should be positive-definite, i.e. $\gamma>0$.  We thus immediately obtain $\beta_{\varphi\sigma}>0$. Moreover, we require $\widetilde{V}_{\rm eff}(C)>\widetilde{V}_{\rm eff}(B)$ in order for [B] to be the global minimum, which yields the condition $\beta_\varphi\beta_\sigma>(2\beta_{\varphi\sigma}-\gamma)^2$, since $\widetilde{V}_{\rm eff}(B)=0$. Owing to this condition, the (true) vacuum manifold of $\widetilde{V}_{\rm eff}$ is always equivalent to $U(1)$ even if the current-carrier exists. 

If the latter condition is not satisfied, namely if $\widetilde{V}_{\rm eff}(A)>\widetilde{V}_{\rm eff}(B)>\widetilde{V}_{\rm eff}(C)$, 
the potential is similar to the one shown in the lower panel of Fig.~\ref{fig:potential}, and the fields $\varphi$ and $\sigma$ tend to go into the global minimum [C]. Roughly speaking, in this case, the interior of the string continues to grow forever since [C] is energetically favoured. As shown in the next subsection, a metastable static string configuration that does not satisfy this latter condition can indeed be constructed numerically. Of course, the configuration of such a metastable string is broken if it collides with another string. 

In summary, in order to obtain a superconducting string configuration, we require
\begin{align}
 2\beta_{\varphi\sigma} > \gamma>0, \quad
 \beta_\varphi\beta_\sigma>(2\beta_{\varphi\sigma}-\gamma)^2.
\label{eq:cond}
\end{align}
These conditions, which are determined only by considering the shape of the effective potential, are necessary but not sufficient for the obtention of a superconducting string. In reality, the configuration is determined by the balance of the potential energy and the gradient energy of the string. In the next subsection, we compute the region of parameter space in which superconducting strings form and are viable by solving \cref{eq:f} to \cref{eq:g} numerically.

%%%%%%%%%%%%%%%%%%%%%%%%%%%%%%%%%%%%%%%%%%%%%%%%%%%%%%%%%%%%%%%%%%%%%%%%%%%%%%%
\subsection{Viable parameter regions for static and stationary strings}
\label{subsec:static}
%%%%%%%%%%%%%%%%%%%%%%%%%%%%%%%%%%%%%%%%%%%%%%%%%%%%%%%%%%%%%%%%%%%%%%%%%%%%%%%
In order to take the gradient energy of the string configuration into account, we solve \cref{eq:f} to \cref{eq:g} numerically.  We impose the following boundary conditions for $f,\alpha$ and $g$, 
\begin{align}
\begin{cases}\displaystyle
f(0)= \alpha(0)= \left.\frac{dg}{dx}\right|_{x=0} = 0, \\
f(\infty)=\alpha(\infty)=1, \quad g(\infty) = 0,
\end{cases}
\label{eq:boundary_inf}
\end{align}
The conditions for $f$ and $\alpha$ are the same as those in the Abelian-Higgs model.
The condition for $g$ stems from the regularity of the field profile at $x=0$ and the requirement that the current vanishes far from the string core.

In practice, we truncate the radial coordinate at finite $x$. The values of $f,\,\alpha$ and $g$ for large $x$ can be computed by considering the asymptotic behaviour of their field equations (see Ref.~\cite{VilenkinShellard} for the Abelian-Higgs case and Ref.~\cite{Peter:1992dw} for the case of superconducting strings). It is easy to find that $\delta f(x)=1-f(x), \delta\alpha(x) = 1-\alpha(x)$, and $g(x)$ satisfy a modified Bessel equation of the second kind in the asymptotic region. More precisely, they behave as $\delta f(x) \approx K_0(\sqrt{2\beta_\varphi}x), \delta \alpha(x) \approx xK_1(\sqrt{2}x)$ and $g(x) \approx K_0(\sqrt{\gamma}x)$ for $\gamma>0$. Therefore we may impose those boundary values for $f(x),\alpha(x)$ and $g(x)$ at $x=x_{\rm max}$ instead of the boundary conditions at infinity. The typical value of $x_{\rm max}$ in our study is $x_{\rm max}=50$, chosen so that $g(x_{\rm max})$ is sufficiently close to zero, say $g(x_{\rm max})\sim 10^{-10}$. As a result, the boundary conditions are
\begin{align}
 \begin{cases}
  f(0) &= 0, \\
  f(x_{\rm max}) &= 1-K_0(\sqrt{2\beta_\varphi}x_{\rm max}), 
 \end{cases}
\quad
 \begin{cases}
  \alpha(0) &= 0, \\
  \alpha(x_{\rm max}) &= 1-x_{\rm max}K_1(\sqrt{2}x_{\rm max}), 
 \end{cases}
\quad
 \begin{cases}
  \frac{dg}{dx}(0) &= 0, \\
  g(x_{\rm max}) &= K_0(\sqrt{\gamma}x_{\rm max}).
 \end{cases}
\label{eq:boundary}
\end{align}
We solve the field equations using an iterative method developed in Ref.~\cite{Hiramatsu:2012xj} with spatial resolution $\Delta x=0.25$.
\begin{figure}[!ht]
\centering{
\includegraphics[width=8.75cm]{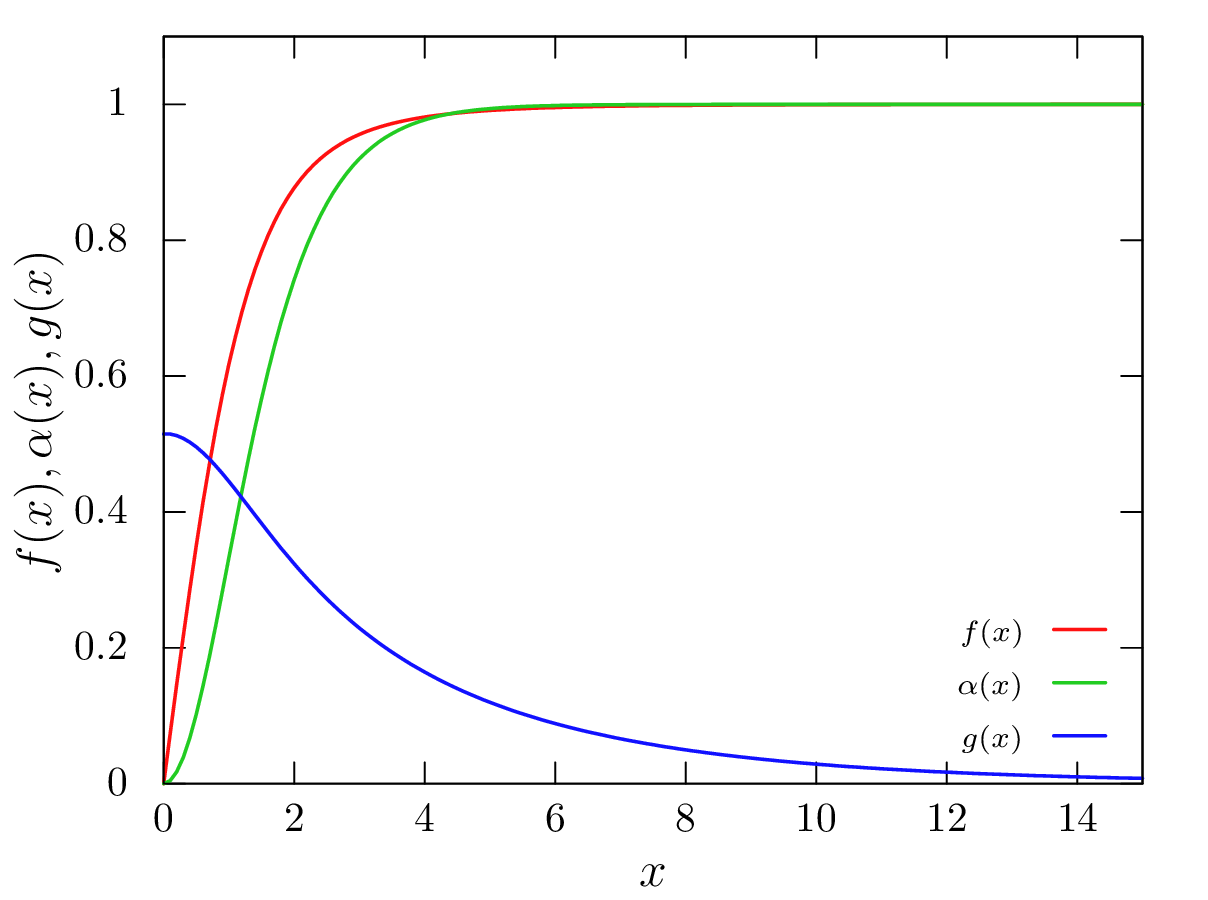}
\includegraphics[width=8.75cm]{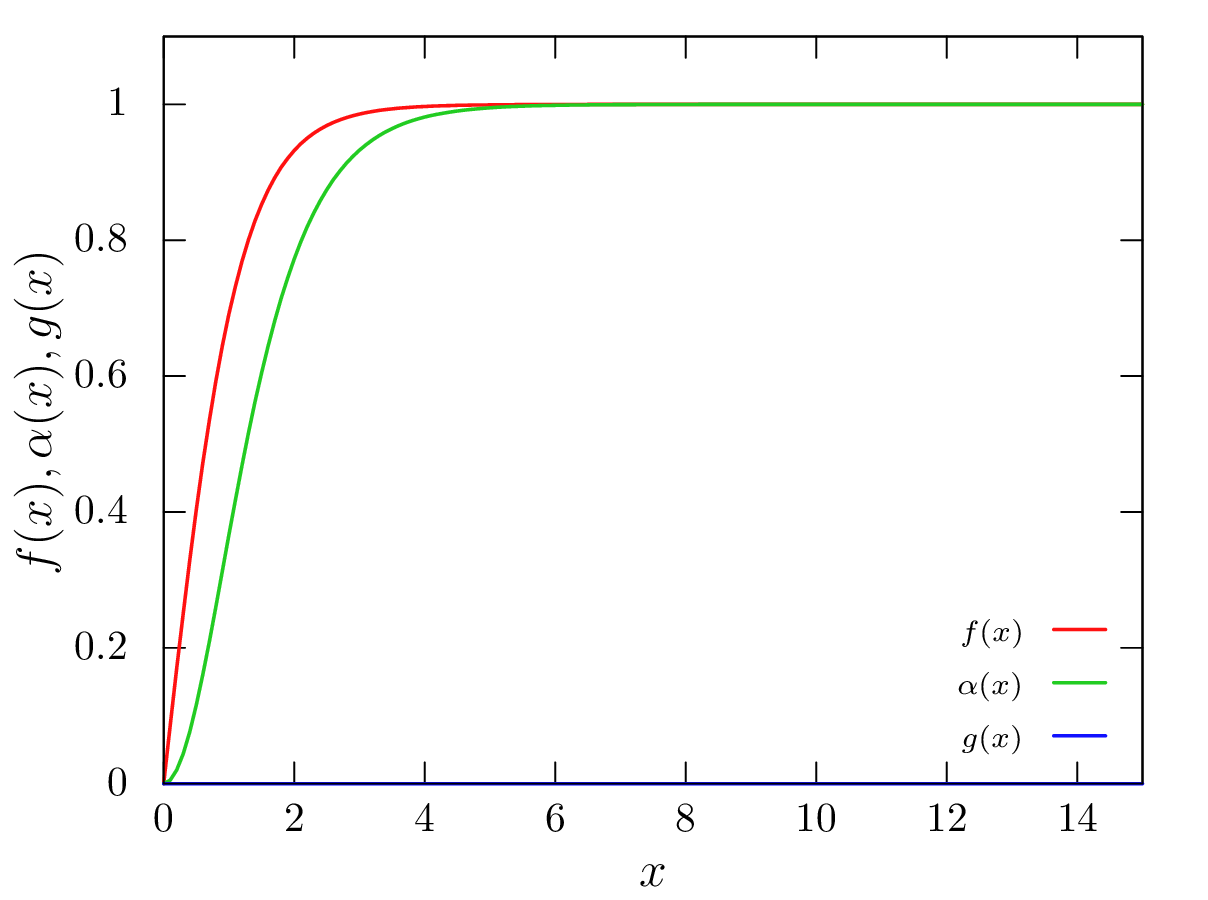}
}
\caption{The field configuration of $f(x)$ (red), $\alpha(x)$ (green) and $g(x)$ (blue) with $\gamma=0.05$, $\beta_\varphi=\beta_\sigma=1$ and $\beta_{\varphi\sigma}=0.49$ (left-hand plot) and $\beta_{\varphi\sigma}=0.30$ (right-hand plot).  Note that $g(x)$ in the right-hand plot is too small to be seen.}
\label{fig:conf2}
\end{figure}
\begin{figure}[!ht]
\centering{\includegraphics[width=12cm]{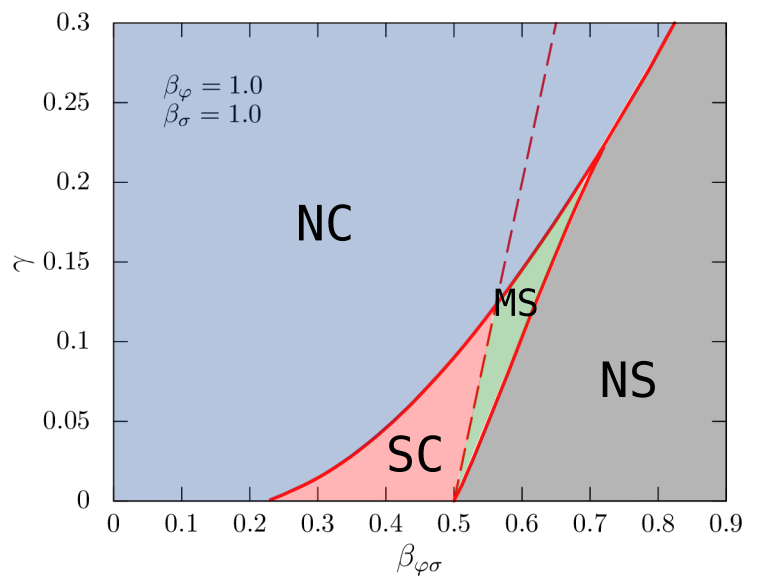}}
\caption{Regions of viability of strings in the $(\beta_{\varphi\sigma}, \gamma)$ plane for $\beta_{\varphi} = \beta_{\sigma} = 1$. The viable parameter region for a superconducting string is the red region labelled ``SC'' corresponding to field profiles as in the left-hand plot of Fig.~(\ref{fig:conf2}). Strings form without a condensate in the blue region labelled ``NC''. The corresponding field profiles are shown in the right-hand plot of Fig.~(\ref{fig:conf2}), with $g(x)$ negligibly small.  No string solutions (NS) exist in the grey region labelled ``NS''. This region is where one cannot obtain plausible field profiles which satisfy the boundary conditions (\ref{eq:boundary}) in the asymptotic region.  Metastable strings (``MS'') form in the small region separated from region ``SC'' by the dashed line (defined by $\gamma=2\beta_{\varphi\sigma}-\sqrt{\beta_\varphi\beta_\sigma}$); in this region, superconducting strings are not stable against fluctuations.}
\label{fig:param}
\end{figure}
The desirable field configurations of $f(x)$, $\alpha(x)$ and $g(x)$
are shown in the left-hand plot of~\cref{fig:conf2}, where $(\beta_{\varphi},\beta_{\sigma},\beta_{\varphi\sigma},\gamma)=(1,1,0.49,0.05)$.
This parameter set yields a vortex solution on which $\sigma \propto g$ takes a non-zero value in the string core.  In contrast, when $\beta_{\varphi\sigma}$ is small, the current-carrier $\sigma$ no longer condenses on the string, as shown in the right-hand plot of~\cref{fig:conf2}, where we set $\beta_{\varphi\sigma}=0.30$.
This string behaves as an Abelian-Higgs string. In this case, the difference between $\widetilde{V}_{\rm eff}(C)$ and $\widetilde{V}_{\rm eff}(A)$  is too small to compensate for the increase in the gradient energy $\sim |\partial_x\sigma|^2$. 

The regions of viability for superconducting strings and non-superconducting strings, obtained by solving Eqs.~(\ref{eq:f})-(\ref{eq:g}), in the range $ 0 \leq \gamma \leq 0.3$ and $0 \leq \beta_{\varphi\sigma} \leq 0.9$, with $\beta_{\varphi}=\beta_{\sigma} = 1$ are shown in Fig.~\ref{fig:param}.  In \cref{appsec:viable} we report on the cases with $\beta_{\sigma}=0.5$ and $1.0$ and $\beta_{\varphi}=0.1,\,0.5$ and $1.0$.

Superconducting strings (with field profiles as in the left-hand plot of Fig.~\ref{fig:conf2}) are obtained in the region labelled ``SC''. This region is separated from a neighboring region labelled ``MS'' in which strings are in a meta-stable state. The line separating the two regions is given by $\gamma=2\beta_{\varphi\sigma}-\sqrt{\beta_\varphi\beta_\sigma}$.  As discussed in~\cref{subsec:condition}, to the right of the dashed line of Fig.~\ref{fig:param}, extremum [B] (see~\cref{fig:potential}) is not the global minimum of the potential. This is why a superconducting string solution in the ``MS'' region is meta-stable.  For small $\beta_{\varphi\sigma}$ or large $\gamma$, strings can form, but no condensate can be obtained (region ``NC'' in Fig.~\ref{fig:param}). The corresponding field profiles are shown in the right-hand plot of Fig.~\ref{fig:conf2} with $g(x)$ suppressed to a level of order $10^{-10}$.  For large $\beta_{\varphi\sigma}$, strings do not form (region ``NS'' in Fig.~\ref{fig:param}). In this region, one cannot obtain field profiles satisfying the boundary conditions (\ref{eq:boundary}) in the asymptotic region. In fact, in this region, the potential energy at extremum [C] in~\cref{fig:potential} becomes large and negative.

The numerical study conducted in this section of the paper, which is restricted to the case of static string configurations, reveals that there exists sharp transitions between the ``NC'', ``SC'' and ``NS'' regions. These transitions are represented by solid red lines.  Note also that our study is not able to resolve the transition between the ``SC'' and ``MS'' regions, which instead stems from the analysis of~\cref{subsec:condition}. The expression for the boundary of the region in which a string forms but no condensate forms (region ``NC''), obtained by considering the gradient energy as well as the potential energy of a superconducting string, is derived in \cref{appsec:NCregion}. 

While the analysis in this subsection was restricted to static strings, in the following section, we perform dynamical simulations of colliding strings starting from the static superconducting string solutions in the ``SC'' region. 

%%%%%%%%%%%%%%%%%%%%%%%%%%%%%%%
%%%%%%%%%%%%%%%%%%%%%%%%%%%%%%%
\section{Simulation setup}
\label{sec:setup}
%%%%%%%%%%%%%%%%%%%%%%%%%%%%%%%
%%%%%%%%%%%%%%%%%%%%%%%%%%%%%%%

\begin{figure}[!ht]
\includegraphics[width=10cm]{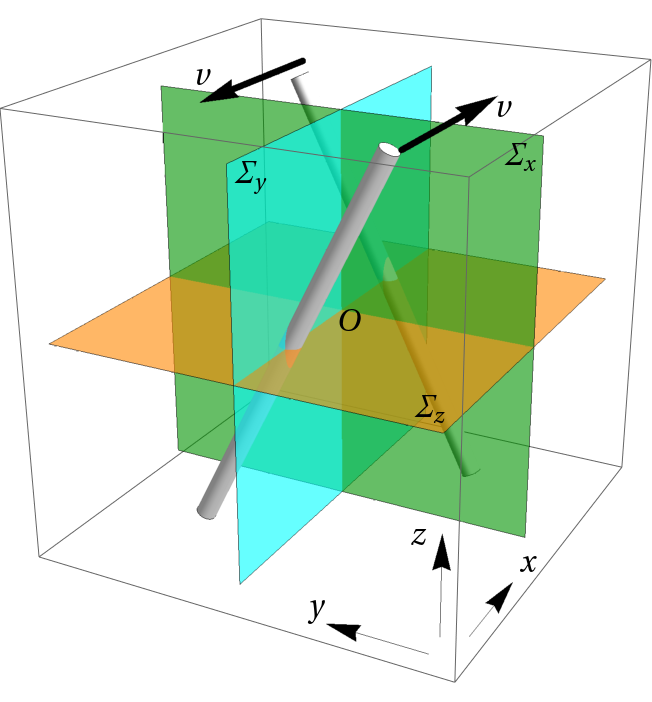}
\caption{Computational domain and initial layout of the strings. $\Sigma_{I}$ for $I=x,y,z$ represents the plane perpendicular to $I$-axis. The strings moves along the $x$-axis with velocity $v$.}
\label{fig:domain}
\end{figure}

\begin{figure}[!ht]
\includegraphics[width=8cm]{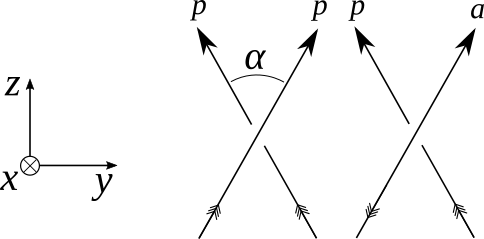}
\caption{The solid arrow represents the direction of $\boldsymbol{B}:=\nabla\times \boldsymbol{A}$, which indicates the winding of a string (the upward arrow represents $n>0$). The triple arrow represents the direction of the conserved current, $\boldsymbol{j}^\sigma$ (the upward arrow represents $K>0$). A string with $\boldsymbol{B} \parallel \boldsymbol{j}^\sigma$ is a {\it parallel} ($p$) string and a string with $\boldsymbol{B} \parallel -\boldsymbol{j}^\sigma$ is an {\it anti-parallel} ($a$) string. We consider two pairs of strings: $(p,p)$ and $(p,a)$.}  
\label{fig:comb}
\end{figure}

Let us now consider the collision of two moving strings.
An individual axisymmetric static vortex solution is obtained by solving \cref{eq:f} to \cref{eq:g} with the boundary conditions given in~\cref{eq:boundary}.  We then obtain a moving string with velocity $v$ and angle $\alpha$ by performing a Lorentz transformation, as explained in~\cref{appsec:moving}. These moving strings are placed in the computational domain as a superposition of the field configurations given in~\cref{eq:multi}. A diagram depicting the initial string configuration is shown in~\cref{fig:domain}. The coordinate origin is at the centre of the domain. It is labelled $O$. The strings have the same initial velocity, $v$, and move along the $x$-axis. For later convenience, we define hyper-surfaces $\Sigma_{I}$ for $I=x,y$ and $z$, as the surface perpendicular to the $I$-axis. The collision angle $\alpha$ is defined as the angle between two colliding strings projected onto the surface $\Sigma_x$ at $x=0$. As seen along the $x$-axis, towards positive values of $x$, the initial configuration and the angle $\alpha$ are shown in \cref{fig:comb}.

We define parallel and anti-parallel strings as the vortex solutions given in \cref{eq:no_phi} to \cref{eq:no_sigma} with ${\rm sign}(n)={\rm sign}(K)$ and ${\rm sign}(n)=-{\rm sign}(K)$, respectively. In other words, the parallel ($p$) string is a string such that the direction of the magnetic flux, $\boldsymbol{B}:=\nabla\times\boldsymbol{A}$, and the direction of the conserved current, $\boldsymbol{j}^\sigma$, are parallel to each other while for the anti-parallel ($a$) string, they are anti-parallel to each other. We can consider two combinations: $(p,p)$ and $(p,a)$, as shown in \cref{fig:comb}. The solid arrow indicates the direction of $\boldsymbol{B}$, and the triple arrow is that of $\boldsymbol{j}^\sigma$, defined in \cref{eq:def_j}. Note that the combination $(a,a)$ is physically equivalent to $(p,p)$, so we do not consider it.

In all simulation runs, we assume $\beta_\varphi=\beta_\sigma=1$ for simplicity. As we mentioned in~\cref{subsec:condition}, stable superconducting strings can be obtained in a restricted region of the $\gamma$--$\beta_{\varphi\sigma}$ plane (recall that $\gamma = K^2-\Omega^2+\mu_\sigma^2$). Hence we can fix $\Omega=0$ without a loss of generality. We then vary other model parameters, $K$, $\mu_\sigma^2$ and $\beta_{\varphi\sigma}$. Then we vary the kinematical parameters of strings, the collision velocity $v$ and the collision angle $\alpha$. 

For each simulation, we solve the dynamical equations \cref{eq:eqphi} to \cref{eq:eqsigma} in Cartesian coordinates. The strings are assumed to extend infinitely outside the computational domain. To realise this, we use \cref{eq:multi} as the boundary condition.

We use the Leap-Frog method in the time domain and approximate the spatial derivatives using second-order central finite differences. For consistency with the boundary conditions, each simulation ends before the result of the collision reaches the boundaries of the computational domain. We compute the optimal size of the computational domain by considering the velocity of strings and the collision angle. For details, see~\cref{appsec:gridsize}. 

Finally, we classify the final configuration of strings after the collision based on the criteria introduced in~\cref{subsec:class}. 

%%%%%%%%%%%%%%%%%%%%%%%%%%%%%%%%%%%%%%%%%%%%%%%%%%%%%%%%%%%%%%%%%%%%%%%%%%%%%%%
%%%%%%%%%%%%%%%%%%%%%%%%%%%%%%%%%%%%%%%%%%%%%%%%%%%%%%%%%%%%%%%%%%%%%%%%%%%%%%%
\section{Final states of colliding strings}
\label{sec:results}
%%%%%%%%%%%%%%%%%%%%%%%%%%%%%%%%%%%%%%%%%%%%%%%%%%%%%%%%%%%%%%%%%%%%%%%%%%%%%%%
%%%%%%%%%%%%%%%%%%%%%%%%%%%%%%%%%%%%%%%%%%%%%%%%%%%%%%%%%%%%%%%%%%%%%%%%%%%%%%%

%\noindent
%[Memo]
%\begin{itemize}
%\item A. Reconnected (p,p) pair (reproduction of Laguna-Matzner's simulations)
%\item A. Reconnected (p,a) pair (disappearance of current and its recovery)
%\item B. Bound state (like Type-I)
%\item C. Double reconnection (like Type-II)
%\item D. Expanding bubble (dynamical instability)
%\item E. phase diagram
%\end{itemize}

%%%%%%%%%%%%%%%%%%%%%%%%%%%%%%%%%%%%%%
\subsection{Regular intercommutation}
%%%%%%%%%%%%%%%%%%%%%%%%%%%%%%%%%%%%%%

\begin{figure}[!ht]% m470b
\begin{minipage}{0.175\textwidth}
\includegraphics[width=3cm,keepaspectratio]{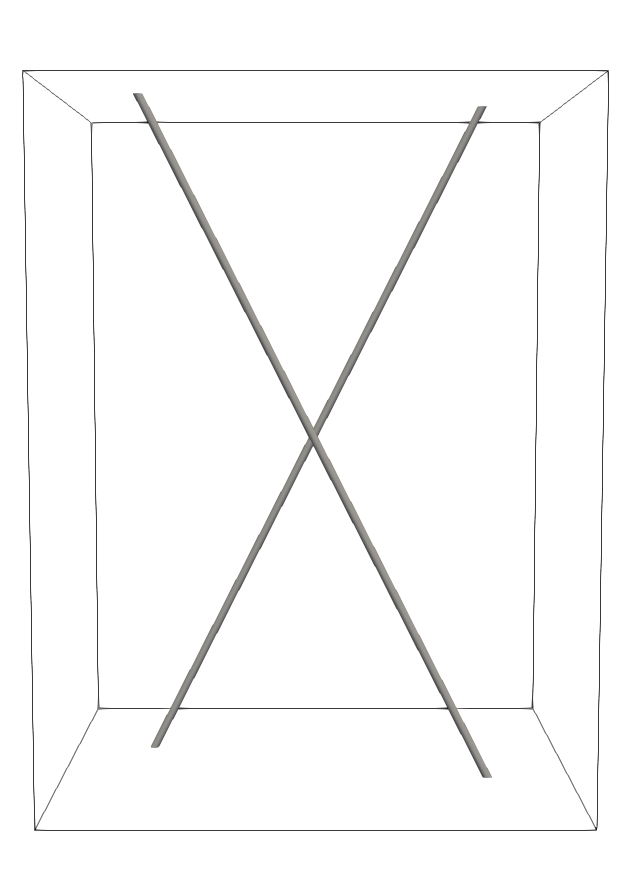}
\begin{center}(a)\end{center}
\end{minipage}
\begin{minipage}{0.175\textwidth}
\includegraphics[width=3cm,keepaspectratio]{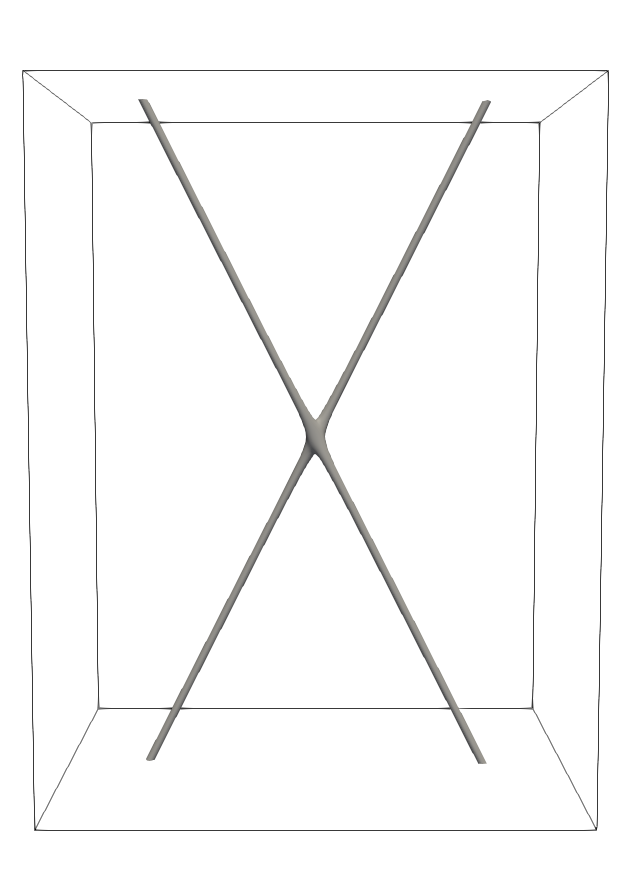}
\begin{center}(b)\end{center}
\end{minipage}
\begin{minipage}{0.175\textwidth}
\includegraphics[width=3cm,keepaspectratio]{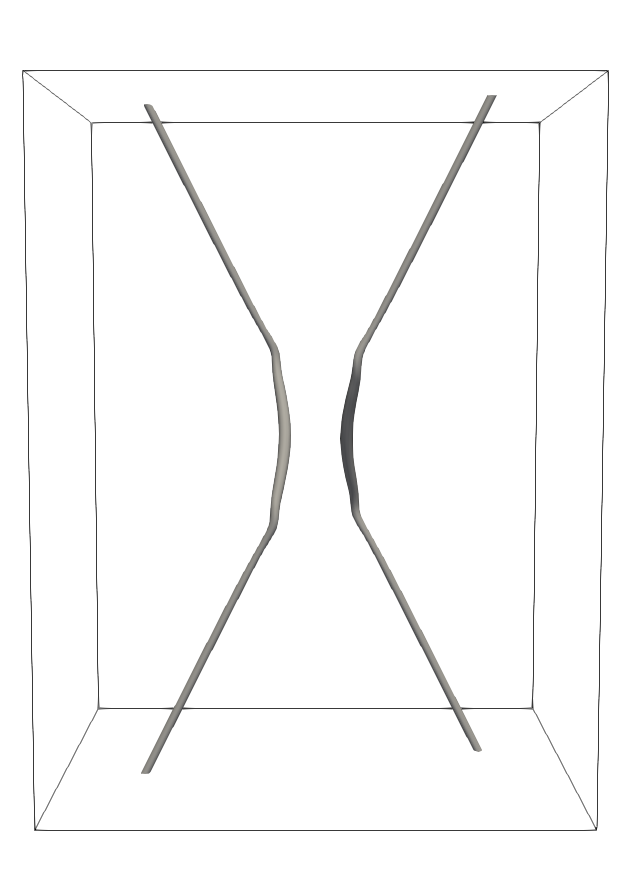}
\begin{center}(c)\end{center}
\end{minipage}
\begin{minipage}{0.175\textwidth}
\includegraphics[width=3cm,keepaspectratio]{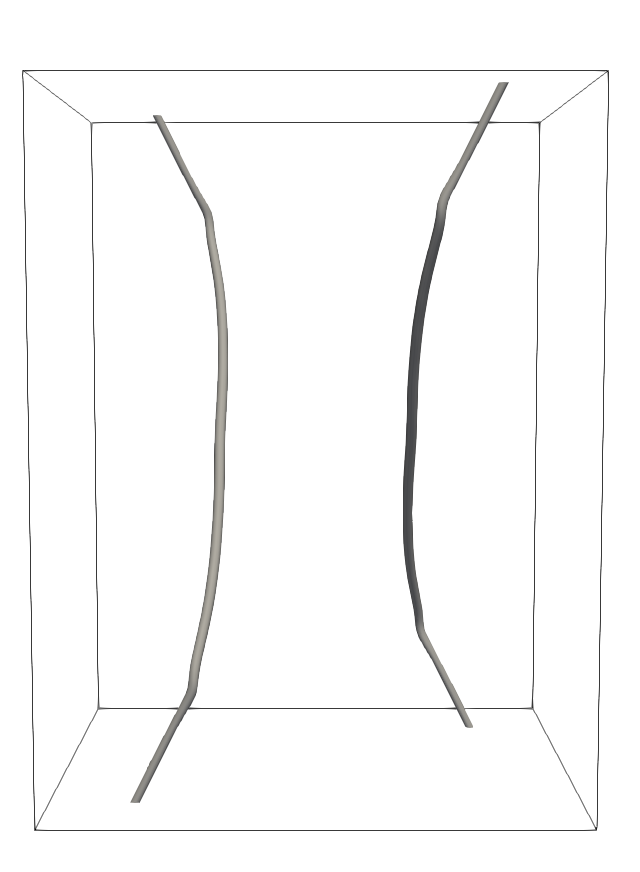}
\begin{center}(d)\end{center}
\end{minipage}
\begin{minipage}{0.175\textwidth}
\includegraphics[width=3cm,keepaspectratio]{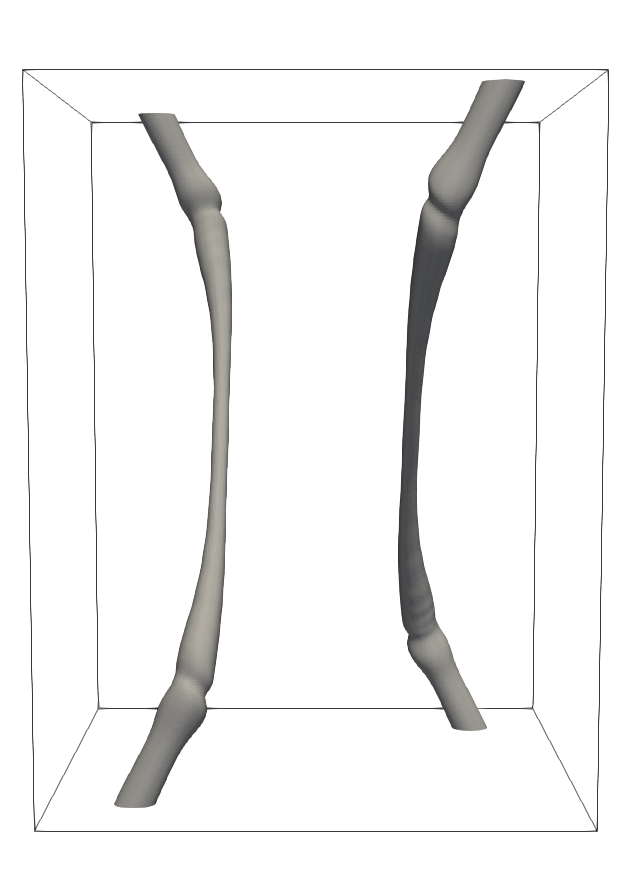}
\begin{center}(e)\end{center}
\end{minipage}
\caption{An example of the reconnection of strings with 
$(\beta_{\varphi\sigma},\mu_\sigma^2,K^2)=(0.49,0.01,0.04)$
 and $v/c=0.5, \alpha=0.2\pi$. The panels (a)--(d) show the isosurface with $|\varphi|=\eta/2$ at $t=0,~17.5/(e\eta),~35.0/(e\eta),~70/(e\eta)$, respectively; (e) the fifth 3D surface plot shows the isosurface with $|j_i^\sigma|=0.01\eta/e$ at $t=70/(e\eta)$. }
\label{fig:rec1}
\end{figure}

Let us first illustrate the familiar intercommutation process. We simulate a $(p,p)$ pair with model parameters $(\beta_{\varphi\sigma},\mu_\sigma^2,K^2)=(0.45,0.01,0.04)$ and $v/c=0.5,\, \alpha=0.2\pi$. In~\cref{fig:rec1}, the isosurface defined by $|\varphi|=\eta/2$ at $t=0,~17.5/(e\eta),~35/(e\eta)$ and $70/(e\eta)$ is shown in the first four 3D surface plots while the isosurface defined by $|j^\sigma_i|=0.01\eta/e$ at $t=70/(e\eta)$ is shown in the fifth 3D surface plot.
Unless otherwise specified, in \cref{fig:rec1} to \cref{fig:bubble}, the $z$-axis points upward in the vertical direction, the $x$-axis is oriented outwards and the $y$-axis points from left to right.

Intercommutation occurs at $t\approx 17.5/(e\eta)$ (in the second panel from the left), after which the strings move away from each other. This process is familiar in the Abelian-Higgs model in the absence of $\sigma$. The fifth plot of~\cref{fig:rec1} shows that the current  $j^{\sigma}_{\mu}$ on the reconnected strings survives the collision process.  We shall discuss the evolution of the current during the collision process in ~\cref{subsec:current}.

%%%%%%%%%%%%%%%%%%%%%%%%%%%
\subsection{Bound states}
%%%%%%%%%%%%%%%%%%%%%%%%%%%

An interesting property of colliding strings is the possible formation of bound states ending at string {\it Y-junctions}. This phenomenon is known to appear in Type-I Abelian-Higgs strings where the gauge coupling is stronger than the self-coupling of the scalar field. In our setup, this corresponds to $\beta_{\varphi}<1$ and $\beta_{\varphi\sigma}=0$ \cite{Bettencourt:1994kf,Salmi:2007ah}. We find that superconducting strings, i.e. strings with $\beta_{\varphi\sigma}>0$, can form bound states even when $\beta_{\varphi}=1$. Indeed, while in the context of the Abelian-Higgs model with $\beta_{\varphi}=1$, no interactions between strings exist because the forces due to the gauge interaction and the scalar interaction balance each other out, in the case of a superconducting string and in the presence of the extra field $\sigma$, the balance of forces is modified, and bound states will form even when $\beta_{\varphi}=1$, as long as we set a relatively small velocity and small collision angle. 

\begin{figure}[!ht]% m445c
\begin{minipage}{0.175\textwidth}
\includegraphics[width=3cm,keepaspectratio]{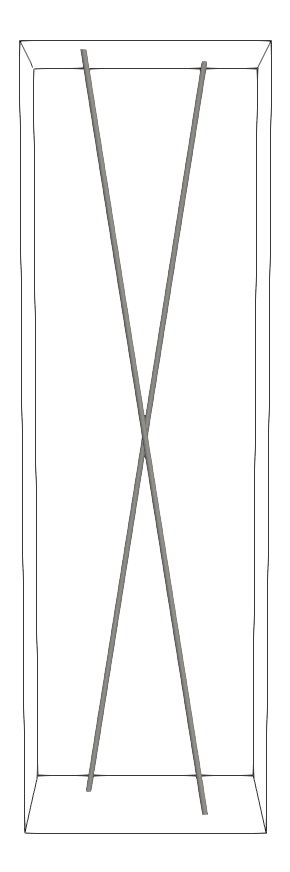}
\begin{center}(a)\end{center}
\end{minipage}
\begin{minipage}{0.175\textwidth}
\includegraphics[width=3cm,keepaspectratio]{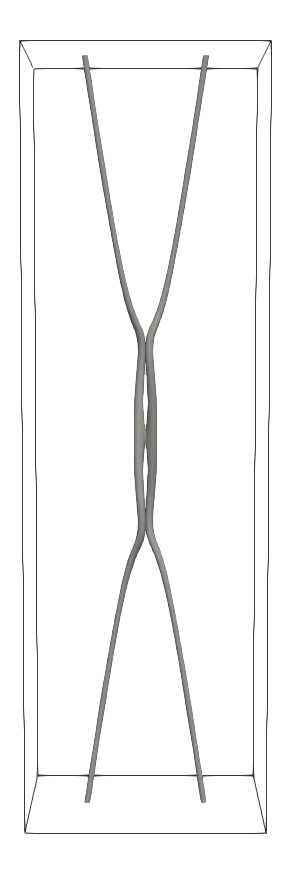}
\begin{center}(b)\end{center}
\end{minipage}
\begin{minipage}{0.175\textwidth}
\includegraphics[width=3cm,keepaspectratio]{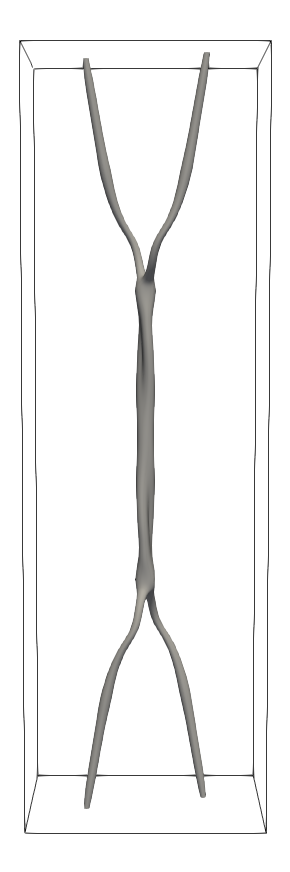}
\begin{center}(c)\end{center}
\end{minipage}
\begin{minipage}{0.175\textwidth}
\includegraphics[width=3cm,keepaspectratio]{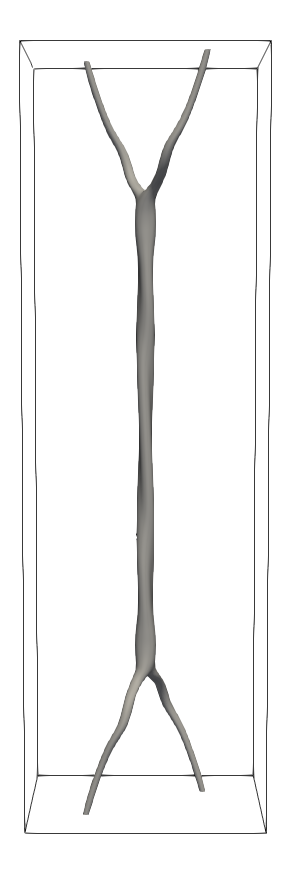}
\begin{center}(d)\end{center}
\end{minipage}
\begin{minipage}{0.175\textwidth}
\includegraphics[width=3cm,keepaspectratio]{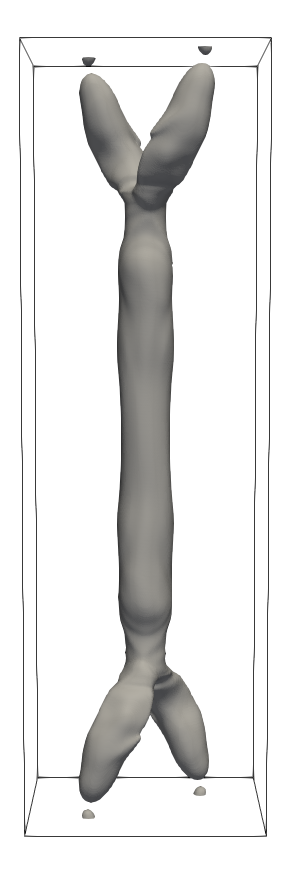}
\begin{center}(e)\end{center}
\end{minipage}
\caption{A bound state is formed after the collision with the same
 model parameters as used in~\cref{fig:rec1}, $(\beta_{\varphi\sigma},\mu_\sigma^2,K^2)=(0.49,0.01,0.04)$, but with different kinematical parameters, $v/c=0.2, \alpha=0.1\pi$. The panels (a)--(d) show the isosurface $|\varphi|=0.5\eta$ at $t=0,50/(e\eta),\,75/(e\eta)$ and $100/(e\eta)$, respectively; (e) shows $|j_i^\sigma|=0.01\eta/e$ at $t=100/(e\eta)$. }
\label{fig:bound}
\end{figure}

An example of a bound state formation for a $(p,p)$ pair is shown in~\cref{fig:bound}. While the model parameters are the same as those in the previous subsection, the kinematic parameters are $v/c=0.2$ and $\alpha=0.1\pi$. The reconnection occurs soon after the collision, but the strong coupling between strings forms the bound state with Y-junctions at $t\approx 75/(e\eta)$. The length of the bound state is expected to grow with time until the string tension interrupts the development of the two Y-junctions. 

%%%%%%%%%%%%%%%%%%%%%%%%%%%%%%%%%%%%%%
\subsection{Double intercommutation}
\label{subsec:double}
%%%%%%%%%%%%%%%%%%%%%%%%%%%%%%%%%%%%%%
In a collision process, strings may intercommute once, as already discussed, or intercommute multiple times \cite{Achucarro:2006es,Achucarro:2010ub}. If an even number of intercommutations occur, the strings will consequently ``pass'' through each other. This phenomenon was first reported for Abelian-Higgs strings with large velocities ($v/c\gtrsim 0.9$) and large collision angles ($\alpha\sim \pi/2$) \cite{Achucarro:2006es}. Here we discuss double intercommutation, as intercommutations occurring more than twice are rare.

\begin{figure}[!ht]
\begin{minipage}{0.175\textwidth}
\includegraphics[width=3cm,keepaspectratio]{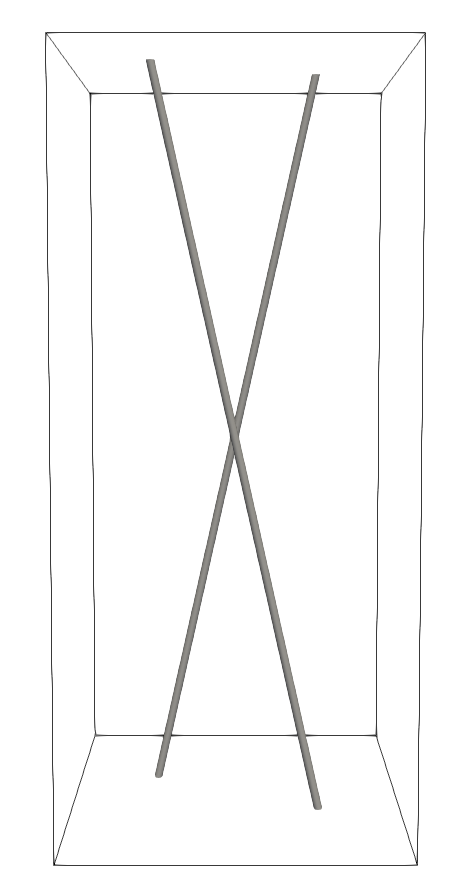}
\begin{center}(b)\end{center}
\end{minipage}
\begin{minipage}{0.175\textwidth}
\includegraphics[width=3cm,keepaspectratio]{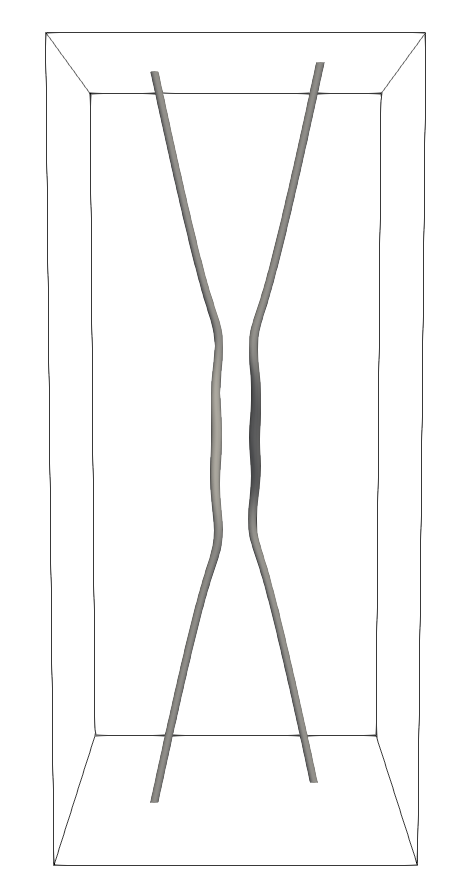}
\begin{center}(b)\end{center}
\end{minipage}
\begin{minipage}{0.175\textwidth}
\includegraphics[width=3cm,keepaspectratio]{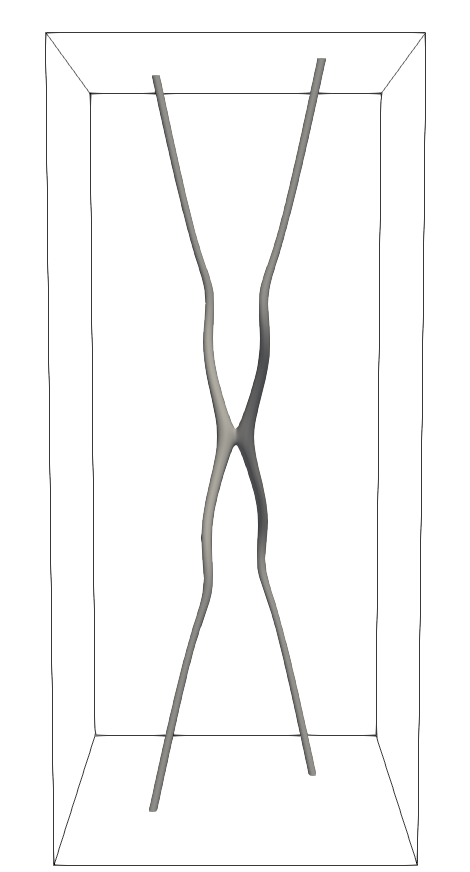}
\begin{center}(c)\end{center}
\end{minipage}
\begin{minipage}{0.175\textwidth}
\includegraphics[width=3cm,keepaspectratio]{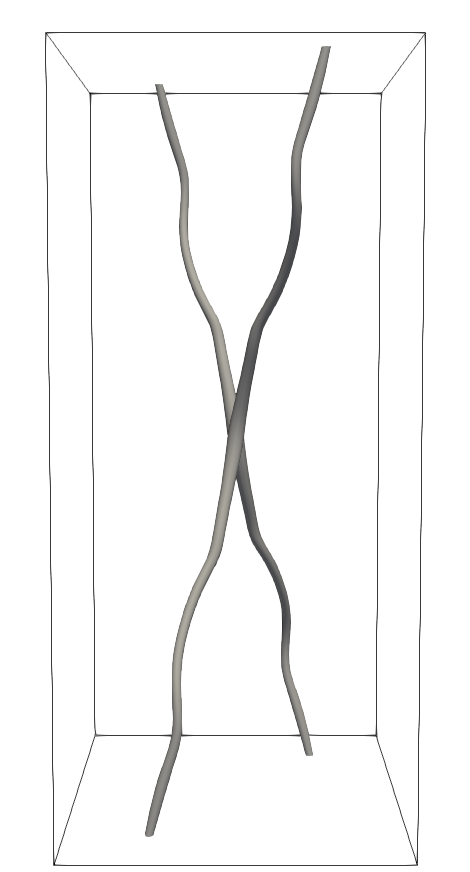}
\begin{center}(d)\end{center}
\end{minipage}
\begin{minipage}{0.175\textwidth}
\includegraphics[width=3cm,keepaspectratio]{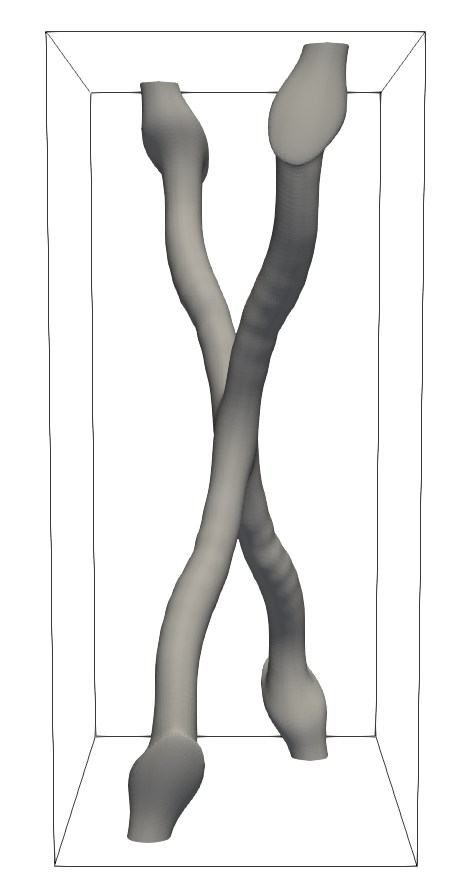}
\begin{center}(e)\end{center}
\end{minipage}
\caption{Double intercommutation process with $(\beta_{\varphi\sigma},\mu_\sigma^2,K^2)=(0.49,0.01,0.04)$
 and $v/c=0.5, \alpha=0.14\pi$. As a result, they slip through each other.
Plots (a)-(d) show the isosurfaces defined by $|\varphi|=\eta/2$ at $t=0, 32.5/(e\eta), 42.5/(e\eta)$ and $70/(e\eta)$, respectively; Plot (e) shows the isosurface  $|j^{\sigma}_i|=0.01\eta/e$ at $t=70/(e\eta)$.}
\label{fig:double}
\end{figure}

In~\cref{fig:double}, we display the case of a $(p,p)$ pair with $(\beta_{\varphi\sigma},\mu_\sigma^2,K^2)=(0.49,0.01,0.04)$, $v/c=0.5$ and $\alpha=0.14\pi$.  In the plots (a)-(d), the 3D surface plots are taken at $t=0,~32.5/(e\eta),~42.5/(e\eta)$ and $70/(e\eta)$ respectively. The strings collide and intercommute in plot (b), similarly to what is shown in plot (c) of~\cref{fig:rec1}. Then, they intercommute once more in plot (c) and ``pass'' through one another; see plot (d). By looking at the locations of the string endpoints at the top and bottom boundaries of the field theory computational domain, one can confirm that the strings intercommute twice and ``pass'' through each other. The currents survive the double intercommutation process, as shown in (e).

%%%%%%%%%%%%%%%%%%%%%%%%%%%%%
\subsection{Expanding bubble}
\label{subsec:expanding}
%%%%%%%%%%%%%%%%%%%%%%%%%%%%%

\begin{figure}[!ht]
\begin{minipage}{0.175\textwidth}
\includegraphics[width=3cm,keepaspectratio]{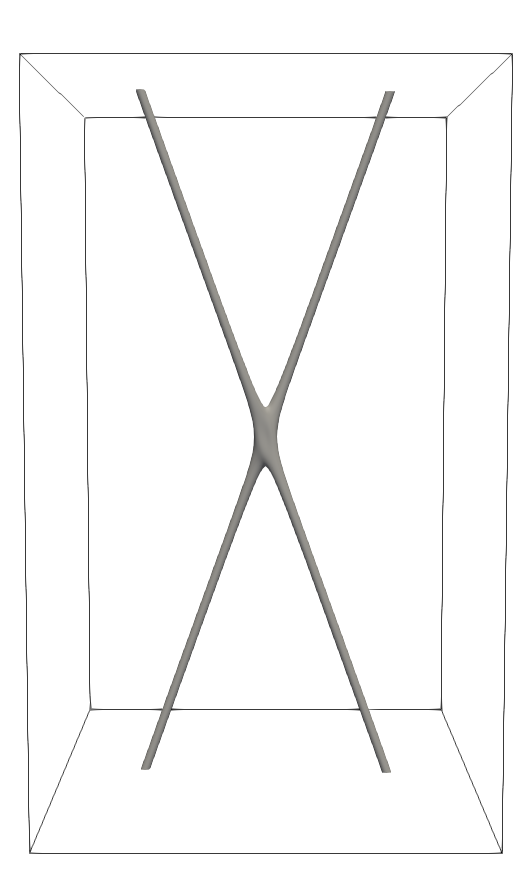}
\begin{center}(a)\end{center}
\end{minipage}
\begin{minipage}{0.175\textwidth}
\includegraphics[width=3cm,keepaspectratio]{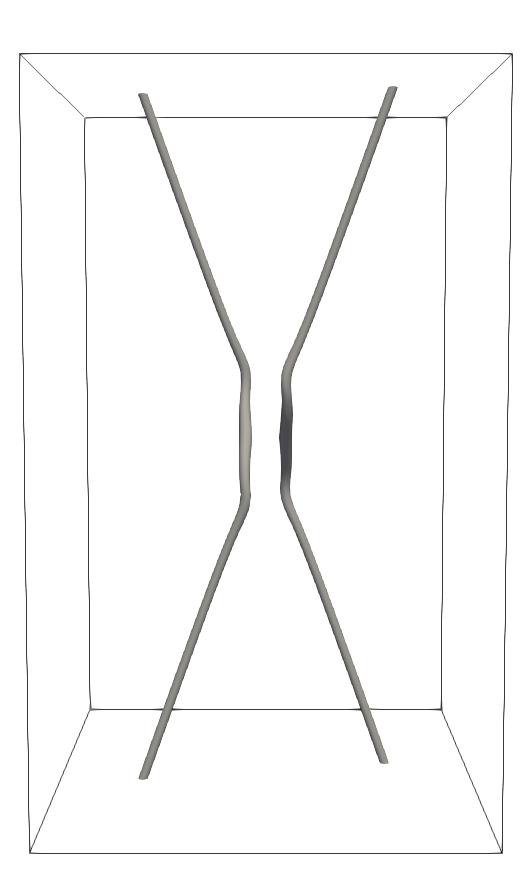}
\begin{center}(b)\end{center}
\end{minipage}
\begin{minipage}{0.175\textwidth}
\includegraphics[width=3cm,keepaspectratio]{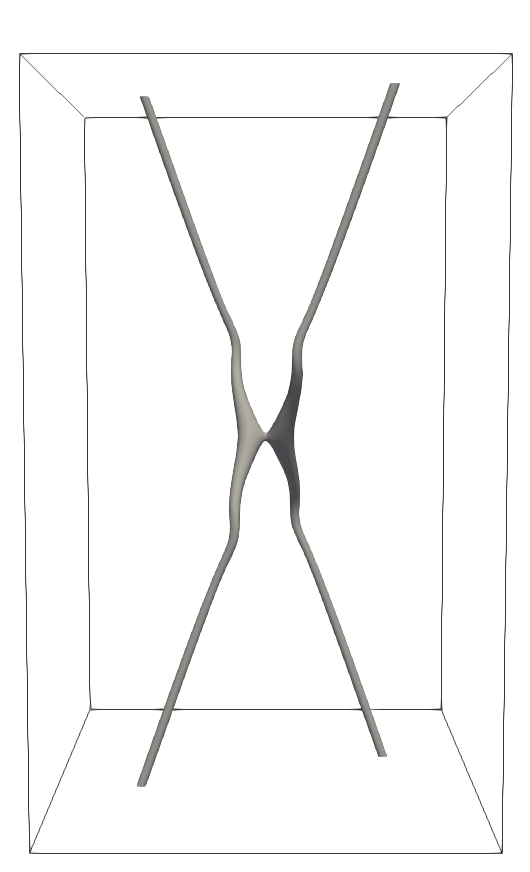}
\begin{center}(c)\end{center}
\end{minipage}
\begin{minipage}{0.175\textwidth}
\includegraphics[width=3cm,keepaspectratio]{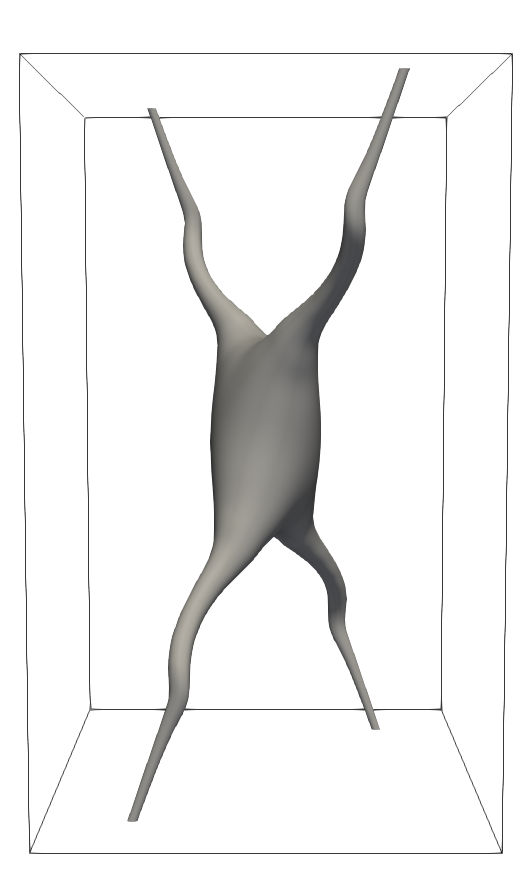}
\begin{center}(d)\end{center}
\end{minipage}
\begin{minipage}{0.175\textwidth}
\includegraphics[width=3cm,keepaspectratio]{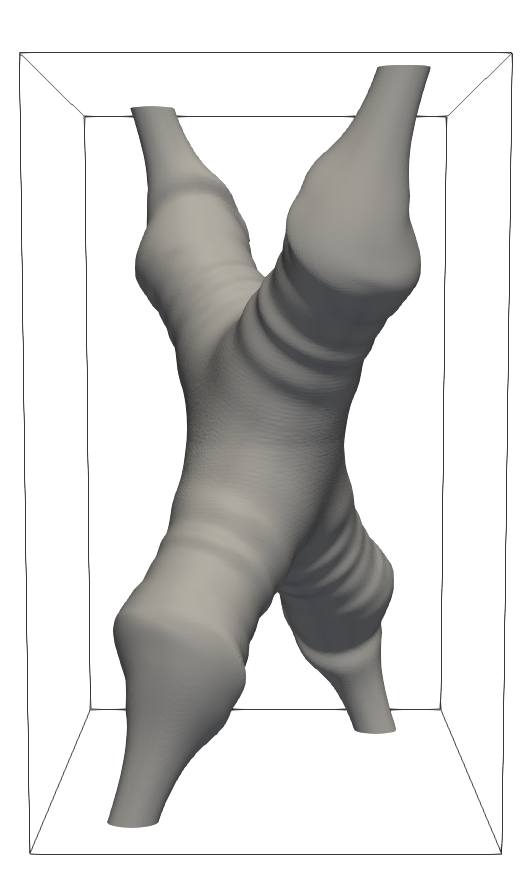}
\begin{center}(e)\end{center}
\end{minipage}
\caption{For $(\beta_{\varphi\sigma},\mu_\sigma^2,K^2)=(0.51,0.01,0.04)$, $v/c=0.6$ and $\alpha=0.22\pi$ the interior of the string expands and then the string configuration breaks down. From left to right, each 3D surface plot shows the
 isosurface $|\varphi|=\eta/2$ at $t=15/(e\eta),~25/(e\eta),~32.5/(e\eta)$ and $65/(e\eta)$ while (e) shows $|j^{\sigma}_i|=0.01\eta/e$ at $t=65/(e\eta)$.}
\label{fig:bubble}
\end{figure}

Finally, superconducting string collisions can result in the nucleation of a bubble. As mentioned in~\cref{subsec:condition} and \cref{subsec:static}, static strings for which the parameters of the effective potential are in the ``MS'' region of~\cref{fig:param} are unstable. Here, we consider a $(p,p)$ pair with $(\beta_{\varphi\sigma},\mu_\sigma^2,K^2)=(0.51,0.01,0.04)$, i.e. $\gamma=0.05$, $v/c=0.6$ and $\alpha=0.22\pi$. Although this set of parameters belongs to the ``SC'' region of~\cref{fig:param}, the dynamical simulation reveals that the string configuration is broken after the collision, as shown in~\cref{fig:bubble}. The strings collide at $t=15/(e\eta)$ in plot (a) and intercommute in plot (b). The reconnected segments then come into contact once more at $t=32.5/(e\eta)$ in plot (c). The reconnected region then expands, see plot (d). In plot (e), we plot the corresponding current amplitude at $t=32.5/(e\eta)$. The currents on the two strings merge into a single current.

As pointed out in~\cref{subsec:condition}, the stable vortex solution cannot be obtained if the vacuum [C] is {\it energetically} favoured (see~\cref{fig:potential}).  Indeed, in such a case, the volume where the fields lie at [C] in the interior of the string grows with time, so that the string configuration is broken, as shown in the right plot of~\cref{fig:bubble}. 
According to~\cref{fig:param}, the case with $\beta_{\varphi\sigma}=0.51$ belongs to the ``NS'' region. Our numerical result suggests that the colliding string has, locally, entered this region.

%%%%%%%%%%%%%%%%%%%%%%%%%%%%%%%%%%%%%%%%%%%%%%%%%%%%%%%
\subsection{Current on strings after collision}
\label{subsec:current}
%%%%%%%%%%%%%%%%%%%%%%%%%%%%%%%%%%%%%%%%%%%%%%%%%%%%%%%
\begin{figure}[!ht]
\includegraphics[width=10.0cm]{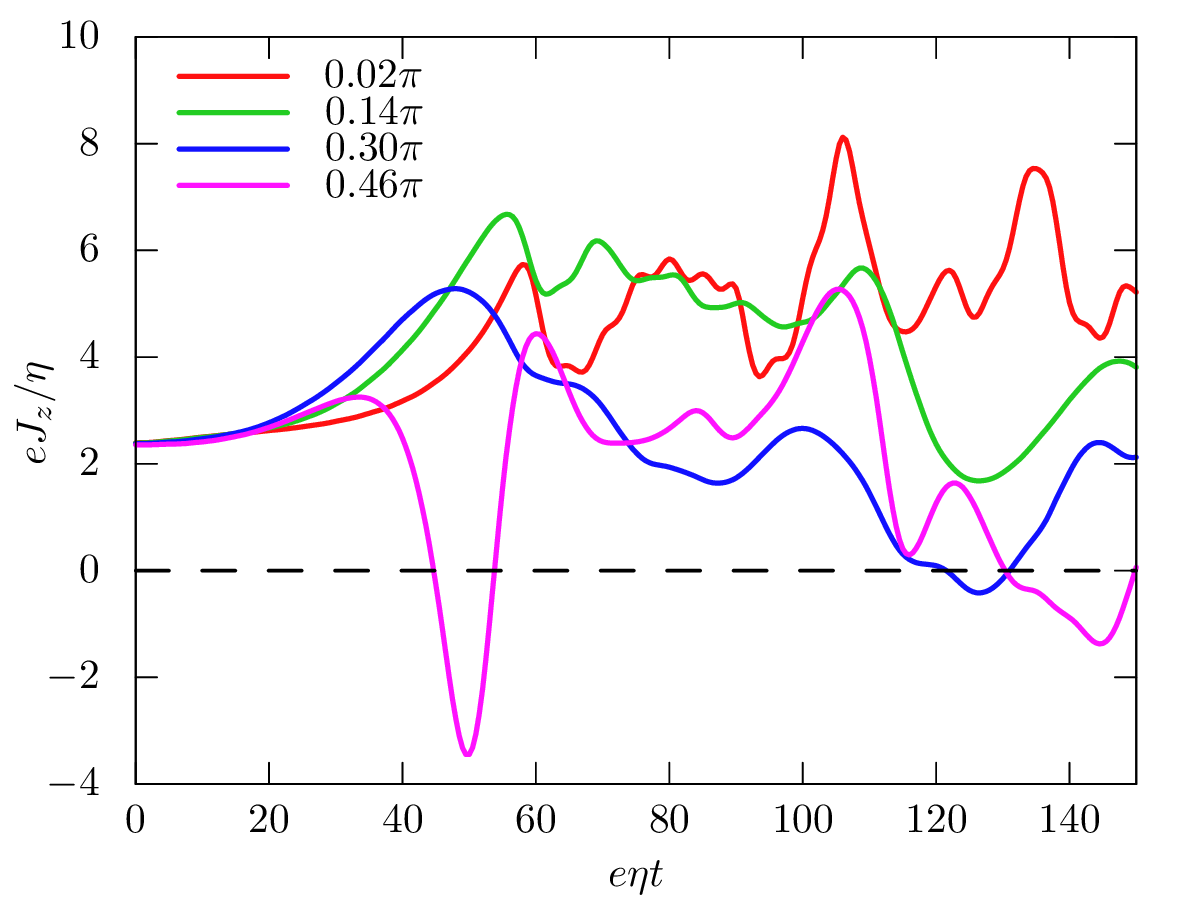}
\caption{The time evolution of the current on $\Sigma_{z}$ 
with $\alpha=0.02\pi$ (red),~$0.14\pi$ (green),~$0.30\pi$ (blue) and $0.46\pi$ (magenta) for a $(p,p)$ pair with $(\beta_{\varphi\sigma},\mu_\sigma^2,K^2)=(0.49, 0.01, 0.04)$ and $v/c=0.1$. 
At the initial time, $J_z(0)\approx 2.4\eta/e$. After the collision, the current on the strings fluctuates significantly. It can sometimes take on negative values. }
\label{fig:time_current}
\end{figure}

To quantify the magnitude of the current of $\sigma$ on the strings after the collision, we define
the current evaluated on the hypersurface $\Sigma_{I}$ for $I=x,y,z$, defined in~\cref{fig:domain}, as
\begin{align}
J_I := \int j^{\sigma}_{\mu}\,d\Sigma^{\mu}_{I}, \label{eq:def_current}
\end{align}
where $j^{\sigma}_{\mu}$ is the conserved current defined in \cref{eq:def_j} and $d\Sigma^{\mu}_{I}$ is the surface element of the hypersurface $\Sigma_{I}$.
For a $(p,p)$ pair, in the case of $(\beta_{\varphi\sigma},\mu_{\sigma}^2,K^2)=(0.49,0.01,0.04)$, the initial currents on the string flow in the positive direction across $\Sigma_{z}$, and its magnitude is $J_z(0) \approx 2.4\eta/e$. There is no current across $\Sigma_{x}$. The currents of each string flow in opposite directions and with the same magnitude across $\Sigma_{y}$, so that the net currents $J_x=J_y=0$.

In~\cref{fig:time_current}, we show the time evolution of the current $J_z(t)$ for $\alpha=0.02\pi$ (red), $0.14\pi$ (green), $0.30\pi$ (blue) and $0.46\pi$ (magenta) with $v/c=0.1$ for a $(p,p)$ pair. The final configurations for $\alpha=0.02\pi, 0.14\pi$ and $0.30\pi$ are bound states, while the choice $\alpha=0.46\pi$ results in regular intercommutation.
We find that, after the collision, the current on the strings fluctuates significantly and can sometimes take on negative values.  We can confirm that finite box size effects do not affect these findings. However, we cannot confirm whether the current will be sustained in the final state at large $t$ in the present numerical setup. To confirm it, one would need a longer simulation with a  larger simulation box.

%%%%%%%%%%%%%%%%%%%%%%%%%%%%%%%%%%%%%%%%%%%%%%%
\section{Classification and phase diagrams}
\label{sec:phase}
%%%%%%%%%%%%%%%%%%%%%%%%%%%%%%%%%%%%%%%%%%%%%%%
%%%%%%%%%%%%%%%%%%%%%%%%%%%%
\subsection{Classification}
\label{subsec:class}
\begin{figure}[!ht]
\begin{minipage}{1\textwidth}

\begin{minipage}{0.19\textwidth}
\includegraphics[width=3cm,height=5cm]{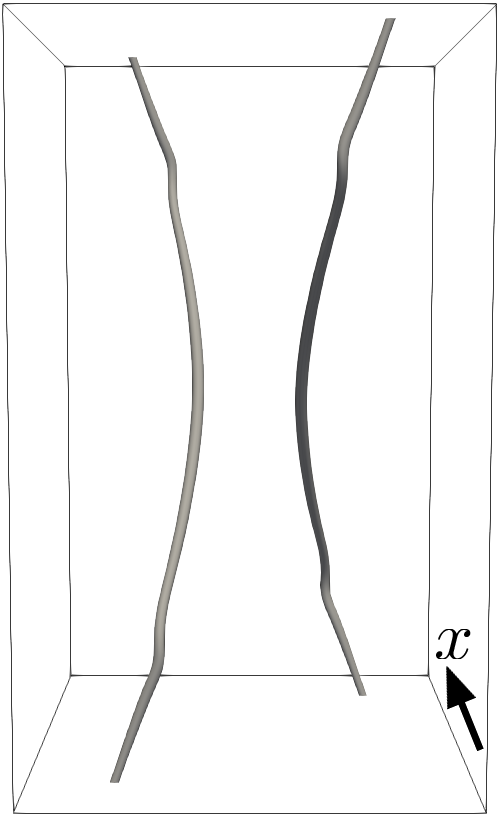}
\end{minipage}
\begin{minipage}{0.19\textwidth}
\includegraphics[width=3cm,height=5cm]{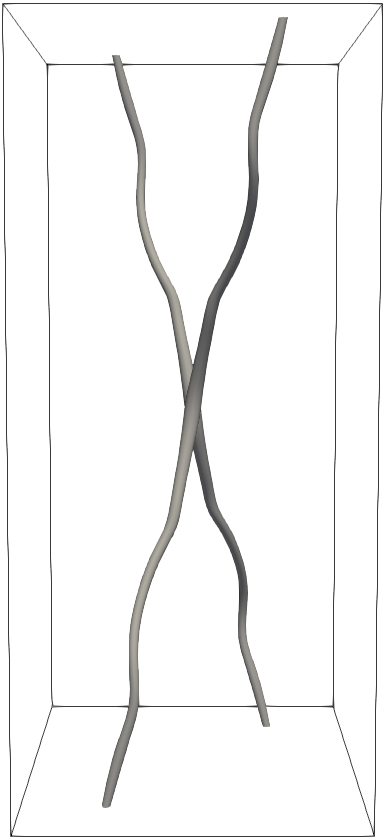}
\end{minipage}
\begin{minipage}{0.19\textwidth}
\includegraphics[width=3cm,height=5cm]{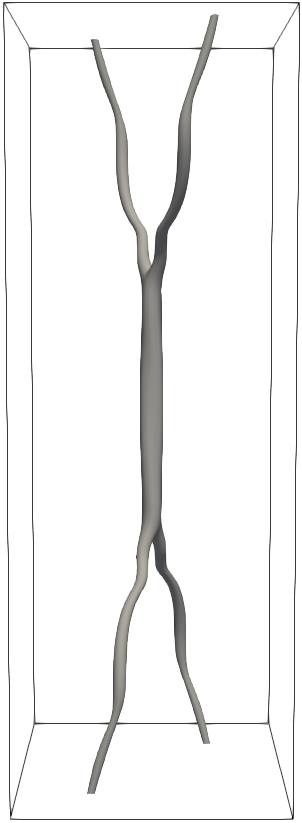}
\end{minipage}
\begin{minipage}{0.19\textwidth}
\includegraphics[width=3cm,height=5cm]{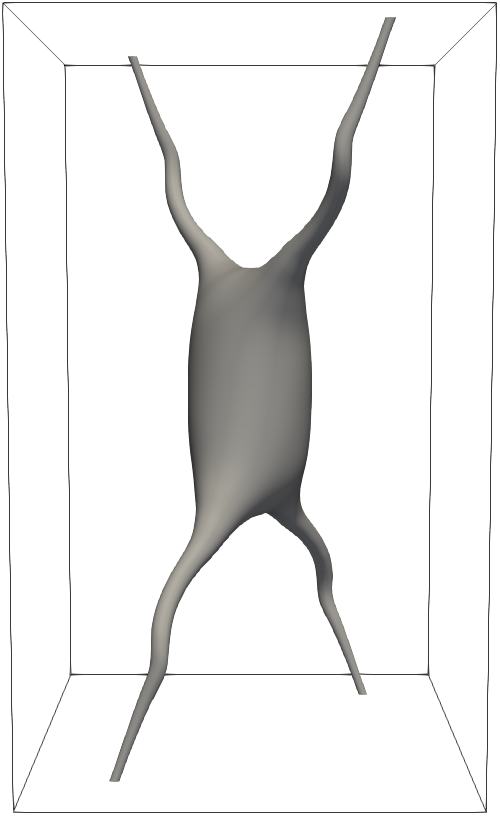}
\end{minipage}
\begin{minipage}{0.19\textwidth}
\includegraphics[width=3cm,height=5cm]{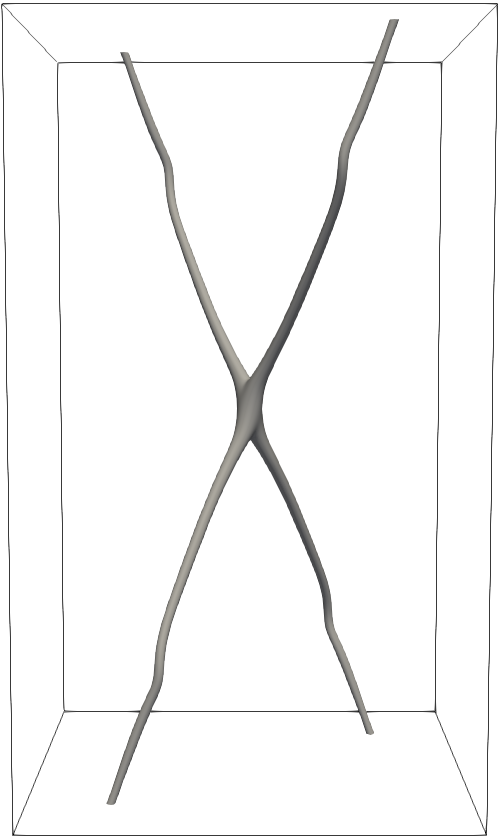}
\end{minipage}
\end{minipage}
\vspace{0.3cm}\\
\begin{minipage}{1\textwidth}
\begin{minipage}{0.19\textwidth}
\includegraphics[width=3cm,height=5cm]{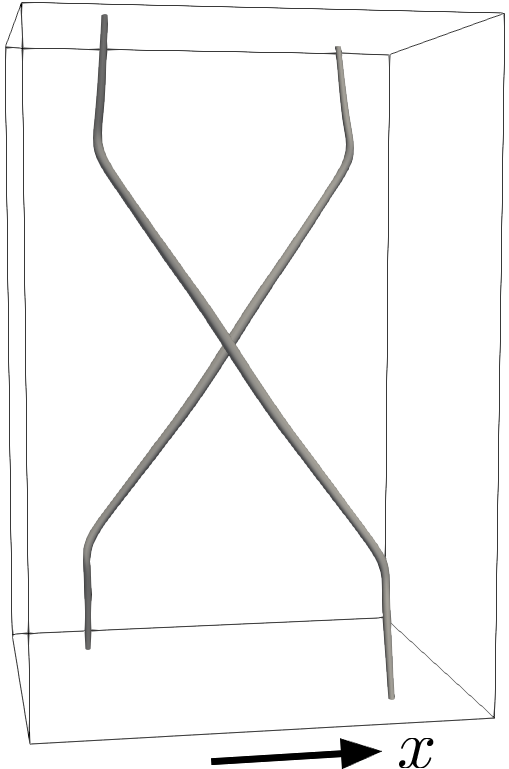}
\begin{center}(a)\end{center}
\end{minipage}
\begin{minipage}{0.19\textwidth}
\includegraphics[width=3cm,height=5cm]{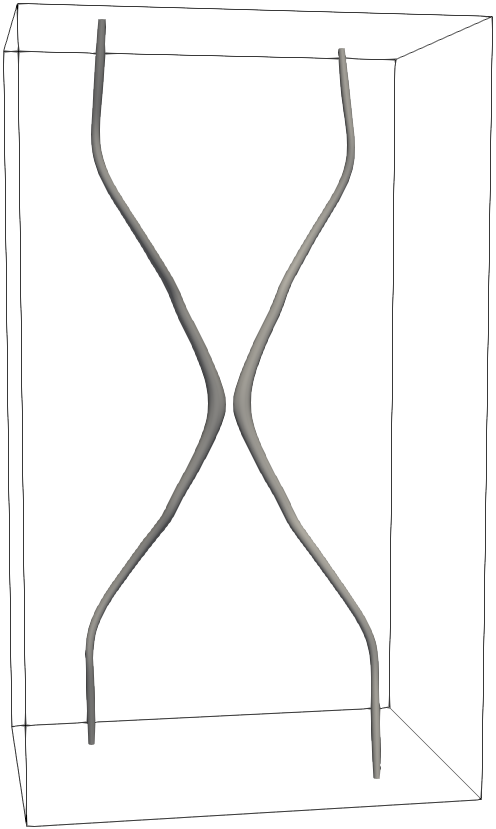}
\begin{center}(b)\end{center}
\end{minipage}
\begin{minipage}{0.19\textwidth}
\includegraphics[width=3cm,height=5cm]{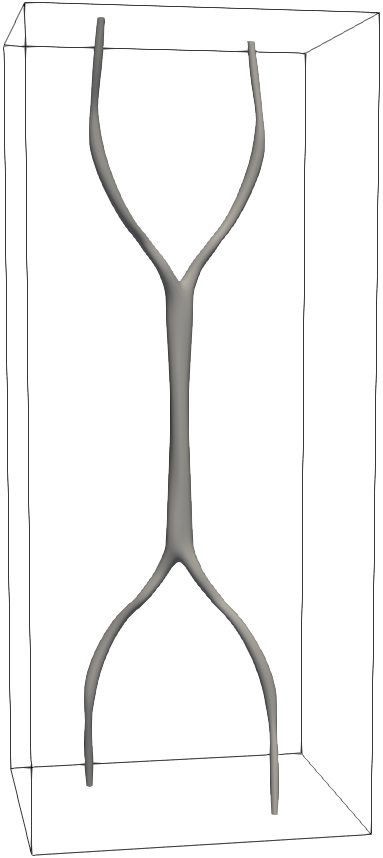}
\begin{center}(c)\end{center}
\end{minipage}
\begin{minipage}{0.19\textwidth}
\includegraphics[width=3cm,height=5cm]{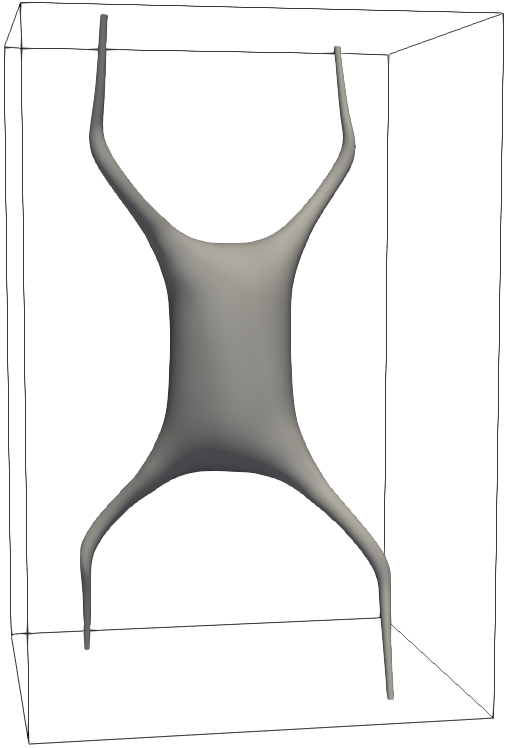}
\begin{center}(d)\end{center}
\end{minipage}
\begin{minipage}{0.19\textwidth}
\includegraphics[width=3cm,height=5cm]{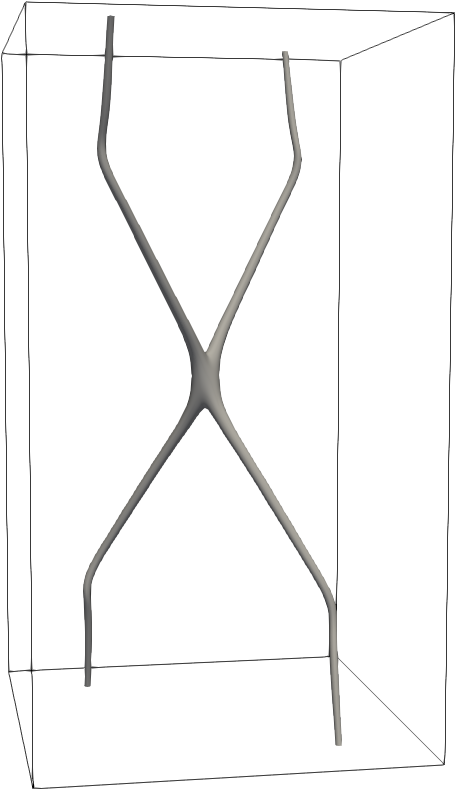}
\begin{center}(e)\end{center}
\end{minipage}
\end{minipage}
\caption{Superconducting string collision outcomes: (a) regular intercommutation, (b) double intercommutation, (c) bound state with Y-junctions, (d) unstable configuration, and (e) indeterminate final state. The lower plots show the upper panels rotated 90 degrees around the $z$-axis, as indicated by the arrows representing the $x$-axis.} 
\label{fig:class}
\end{figure}

%%%%%%%%%%%%%%%%%%%%%%%%%%%%
There is a total of five possible outcomes for superconducting string collisions. A summary of these final states is shown in~\cref{fig:class}. The first four states, (a) to (d), were discussed in previous sections. The last one, state (e), is one for which the outcome of the collision is an indeterminate final string state, which, for longer simulation times, may evolve into either (c) or (d).
 
Let us now introduce some criteria with which we can classify the final states of colliding strings, using the simulated end states of $\varphi(t_{\rm end},\xx)$ and $\sigma(t_{\rm end},\xx)$, at $t_{\rm end}$.  Our code computes, at $t_{\rm end}$, the surface areas for which $|\varphi|<\eta/2$ on $\Sigma_x$, $\Sigma_y$ and $\Sigma_z$, see \cref{fig:domain}. Those surface areas are denoted by $S_x, S_y$ and $S_z$, respectively\footnote{The coordinate origin is located at the centre of the computational box. We set up the simulation so that the collision occurs at that particular location.}.  If the strings undergo regular intercommutation, $S_x \ne 0, S_y=0$ and $S_z\ne 0$ for all $\alpha$. Indeed, since the strings have velocities along the $x$-axis and momentum is conserved after the collision, the strings must cross $\Sigma_x$, while it is the intercommutation process that results in $S_y=0$ and $S_z\ne 0$ for this particular configuration of string collision.  This is confirmed by Fig.~\ref{fig:rec1}.

If the strings undergo double intercommutation, they ``pass'' through each other such that the final string configuration is identical to the starting configuration. In this case, $S_x=0$, see plot (b) in Fig.~\ref{fig:class}.

If $S_x, S_y$ and $S_z$ are all non-zero, the strings either create a bound state or become unstable and the volume for $|\varphi|<\eta/2$ grows with time.  If the strings form a bound state, the final configuration is parallel to the $z$-axis. Hence $S_z$ should be much smaller than $S_x$ and $S_y$ (plot (c) in~\cref{fig:class}). Instead, if $S_x, S_y$ and $S_z$ are comparable, the final state would be like plot (d) or (e) in \cref{fig:class} and thus corresponds to bubble nucleation or an indeterminate string state.

To distinguish between (c) and (d) or (e), we introduce an ellipticity parameter defined as $e_z \equiv S_z^{-1}/(S_x^{-2}+S_y^{-2}+S_z^{-2})^{1/2}$. If the strings are bounded along the $z$-axis, $e_z\sim 1$. In what follows, we consider the final string state to be a bound state when $e_z \geq e_{z,{\rm crit}}=0.9$. 

As mentioned before, because simulation time is limited, some end configurations at $t_{\rm end}$ see plot (e) of~\cref{fig:class} are not in their definitive state and could evolve into either (c) or (d) for longer simulation times. These strings are in an indeterminate state at $t_{\rm end}$. To decide whether a final string configuration is in state (d) or (e), we introduce the effective volume $V = \sqrt{S_xS_yS_z}$, and consider final states with $e_z < e_{z,{\rm crit}}$ and $V\geq V_{\rm crit}=3\times 10^{4}\eta^{-3}$ to be expanding bubbles, while if $V<V_{\rm crit}$, we consider them to be indeterminate string states.

The classification of the final states is summarized \cref{tab:criterion}.
In the following, we use, throughout, the critical values $e_{z,{\rm crit}}=0.9$ and $V_{\rm crit}=3\times 10^{4}\eta^{-3}$ mentioned above.  These criteria, while they have no rigorous physical motivation, are useful for the classification of string end states into types (a) to (e). We confirmed that slight changes in these parameters do not significantly affect our findings.

We do not discuss whether currents remain on the string after the collision. In the previous section, we saw cases for which the current disappears after the collision, but we believe this phenomenon to be transient, with a non-zero current re-established for a simulation with increased duration and a larger computational domain.  While, in principle, the final state of the current is of interest, the current on the strings after the collision continues to fluctuate significantly in our simulations, see \cref{fig:time_current}.  For this reason, in what follows, we do not report on the current's final configuration.  Finally, we would like to point out that bound states may also be transient and may disappear under the effect of the string tension for longer simulation times, particularly for cases in which the bound state is formed on the edge of its region of viability in parameter space. 

\begin{table}[h]
\def\arraystretch{2}%
{\setlength{\tabcolsep}{1.5em}
\begin{tabular}{l|l|l|l}
Criteria & Classification & Colour & Critical values\\
\hline
\hline
$S_x\ne 0$, $S_y=0$ and $S_z\ne 0$   & regular intercommutation & cyan & \hspace{1cm}-- \\
$S_x=0$                              & double intercommutation  & magenta & \hspace{1cm}-- \\
$e_z \geq e_{z,\mathrm{crit}}$       & bound state & orange & $e_{z,{\rm crit}}=0.9$ \\
$e_z < e_{z,\mathrm{crit}}$ and $V >  V_{\mathrm{crit}}$ & expanding bubble & grey & $V_{\rm crit}=3\times 10^{4}\eta^{-3}$ \\
none of the above & indeterminate state & white & \hspace{1cm}-- \\
\hline\hline
\end{tabular}}
\caption{Criteria used in the classification of string collision final states. The column entitled ``Colour'' gives the colour codes used in Fig.~\ref{fig:phase1}-\ref{fig:phase6}.}
\label{tab:criterion}
\end{table}

%%%%%%%%%%%%%%%%%%%%%%%%%%
\subsection{Phase diagram}
\label{subsec:diagram}
%%%%%%%%%%%%%%%%%%%%%%%%%%

In this section, we report on the final states resulting from the collision of superconducting strings in the $(v,\alpha)$ plane. We consider six sets of model parameters, listed in~\cref{tab:param}. 
In Model I, the parameters are the same as the ones used in~\cref{subsec:double}. In Model II, $\beta_{\varphi\sigma}$ is taken to be slightly larger than in Model I. In Models III and IV, we set $K^2$ to be respectively smaller and larger than in Models I and II. In Model V, $\beta_{\varphi\sigma}$ is smaller than in Model III. Finally, in model VI, $K^2$ and $\mu_{\sigma}^2$ are interchanged with those in Model IV, while keeping $\gamma=0.08$. 

We vary the collision velocity $v/c$ from $0.1$ to $0.9$ in steps $\Delta(v/c)=0.05$, and the collision angle $\alpha$ (see~\cref{fig:comb} for the definition of $\alpha$) from $0.02\,\pi$ to $0.46\,\pi$ in steps $\Delta \alpha=0.04\,\pi$. In Models I and II, we also consider angles $\alpha$ in the range $0.58\pi\leq \alpha\leq 0.98\pi$, with $\Delta \alpha = 0.08\,\pi$. We classify the final configurations according to~\cref{tab:criterion}.

%%%%%%%%%%%%%%%%%%%%%%%%%
\subsubsection{Model I}
\label{subsubsec:model1}
%%%%%%%%%%%%%%%%%%%%%%%%%

\begin{figure}[!ht]
\includegraphics[width=8cm]{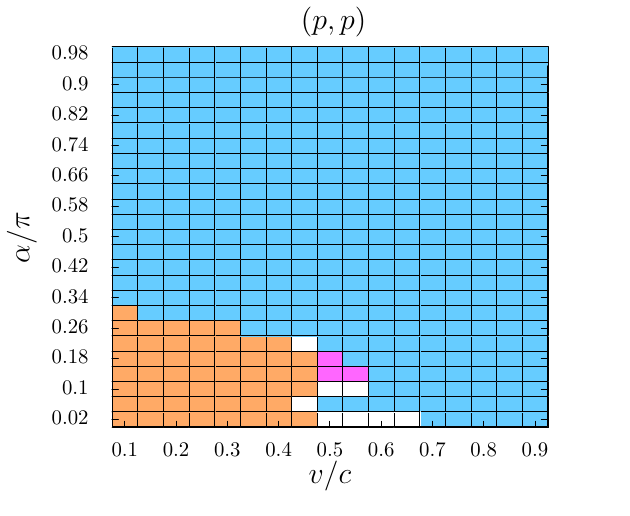}
\includegraphics[width=8cm]{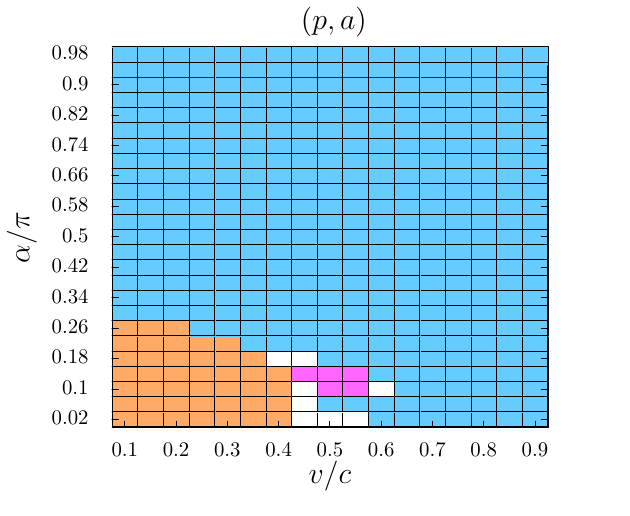}
\caption{The phase diagram of the colliding strings for Model I, with $(\beta_{\varphi},\beta_{\varphi\sigma},\beta_{\sigma},\mu_{\sigma}^2,K^2)=(1.0,0.49,1.0,0.01,0.04)$. The left panel is $(p,p)$ and the right one is $(p,a)$. For the correspondence between the colours in the plots and the string collision outcomes, see Table~\ref{tab:criterion}.}
\label{fig:phase1}
\end{figure}

The phase diagram for Model I is shown in~\cref{fig:phase1} for $(p,p)$ and $(p,a)$-type string pairs (see~\cref{sec:setup} for details).  Strings with $v/c \lesssim 0.5$ and $\alpha \lesssim 0.3\,\pi$ form bound states (orange region). This is similar to the outcome of the collisions between Type-I Abelian-Higgs strings with $e^2>\lambda/2$ \cite{Salmi:2007ah}. Strings with $v/c\sim 0.5$ and $\alpha \sim 0.2\pi$ pass through each other via double intercommutation (magenta region). This behaviour is observed in Type-I/II Abelian-Higgs strings only for strings with large velocities \cite{Achucarro:2006es,Achucarro:2010ub}.  While bound states form in a wide range of velocities and angles, double intercommutation occurs only in a restricted region of the kinematic parameter space.

As is well known, in a bound state, Y-junctions may form~\cite{Avgoustidis:2014rqa,Bevis:2008hg,Copeland:2003bj,Jackson:2004zg,Polchinski:1988cn,Binetruy:2010bq,Matsui:2020hzi,Steer:2017xgh}. At such junctions, the string tension in the two outer strings may unbind the bound string state. Whether this unbinding is observed or not depends on the duration of the simulation. For this reason, the border between the regions of regular intercommutation (cyan region) and bound states (orange region) comprises some indeterminate string states (white region).
\begin{table}[!ht]
\bgroup
\def\arraystretch{2}%
{\setlength{\tabcolsep}{1.5em}
\begin{tabular}{c|ccccccc}
Model  & I & II & III & IV & V & VI \\
%       1+2 6+7    3    4    8   5
\hline
\hline
$\beta_{\varphi\sigma}$ & 0.49 & 0.51 & 0.49 & 0.49 & 0.40 & 0.49 \\ %& 0.99  \\
$\mu_\sigma^2$          & 0.01 & 0.01 & 0.01 & 0.01 & 0.01 & 0.07 \\ %& 0.01  \\
$K^2$                   & 0.04 & 0.04 & 0.02 & 0.07 & 0.02 & 0.01 \\ %& 0.04  \\
$\gamma$                & 0.05 & 0.05 & 0.03 & 0.08 & 0.03 & 0.08 \\ 
\hline
String type in~\cref{fig:param} & SC & MS & SC & SC & SC & SC \\
\hline
\hline
\end{tabular}}
\egroup
\caption{Model parameters. Throughout the simulations, we fix $\beta_\varphi=\beta_\sigma=1$ and $\Omega=0$. 
Note that while the phase diagrams obtained are not sensitive to the value of $\Omega$, they are sensitive to the difference $K^2 - \Omega^2$.}
\label{tab:param}
\end{table}

One also finds that collisions with large angles, $\alpha>\pi/2$, result in regular intercommutation for all values of the velocity (cyan region). If the collision angle is large, the curvature of the outgoing strings around the impact point is significant, and the string velocity near the impact point is large. For this reason, in such a configuration, no bound state can be formed, and no double intercommutation can occur.

\begin{figure}[!ht]
\includegraphics[width=8cm]{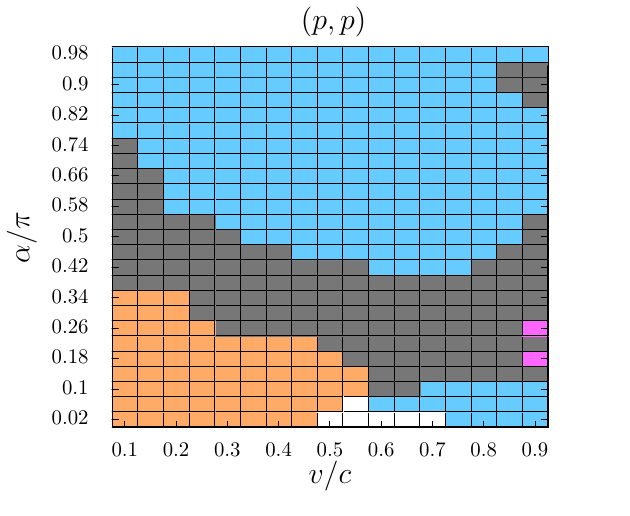}
\includegraphics[width=8cm]{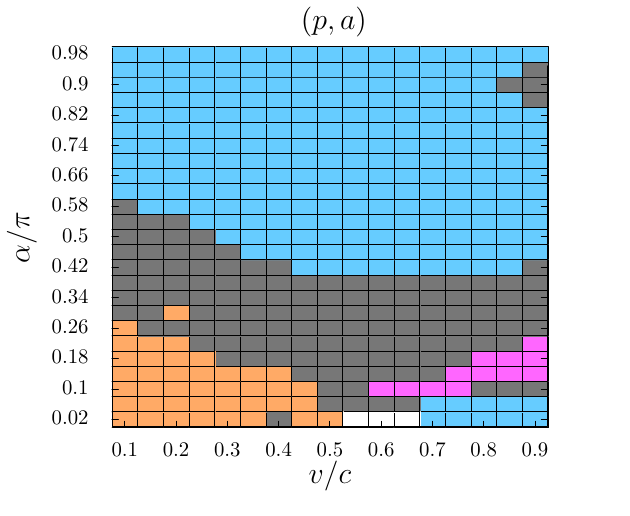}
\caption{Phase diagram of the colliding strings for Model II, with $(\beta_{\varphi},\beta_{\varphi\sigma},\beta_{\sigma},\mu_{\sigma}^2,K^2)=(1.0,0.51,1.0,0.01,0.04)$. The left and right-hand plots correspond to $(p,p)$ and $(p,a)$-type string pairs, respectively. For the correspondence between the colours in the plots and the string collision outcomes, see Table~\ref{tab:criterion}.}
\label{fig:phase2}
\end{figure}

%%%%%%%%%%%%%%%%%%%%%%%%%
\subsubsection{Model II}
\label{subsubsec:model2}
%%%%%%%%%%%%%%%%%%%%%%%%%

Let us now consider the collisions of strings whose parameters belong to the region of Fig.~\ref{fig:param} labelled ``MS''.  In this region, the extremum [C] in~\cref{fig:potential} becomes the global minimum (see lower plot).  Strings in this region of parameter space are dynamically unstable, which means that collisions of such strings can lead to expanding bubbles around the impact point (see~\cref{subsec:expanding}).  This is confirmed by the results of the dynamical simulations, see~\cref{fig:phase2}, which demonstrate that the collisions of strings belonging to model II form bubbles in a large part of the kinematic phase space when $\alpha \lesssim \pi/2$, see~\cref{fig:phase2} while they intercommute when $\alpha \gtrsim \pi/2$.\\

As can be seen from \cref{fig:param}, strings with $\gamma \geq \min\left( 0, 2\beta_{\varphi\sigma} - \sqrt{\beta_{\varphi}\beta_{\sigma}}\right)$ for values $\beta_{\varphi\sigma} \leq 0.5$ belong to region ``SC''.  In a hypothetical bound state resulting from a collision of two model II strings, the total current is given by $2K\cos(\alpha/2)$. Given the constraint on $\gamma$ given above and its definition ($\gamma = K^2+\mu_{\sigma}^2$), one can deduce that the collision of strings in region ``MS'' can result in stable bound states for small angles $\alpha$. In addition, the interaction time will scale as the ratio of the string thickness, $d$, and the velocity of the incident strings, taken to be $v$. That is to say, the interaction time and the probability to form bound states will be greater for small velocities.  

It is worth noting that the region in the $(v/c, \alpha/\pi)$ plane in which the result of the collision of model II strings is an unstable string state is greater for greater values of $\beta_{\varphi\sigma}$.  This can be seen in \cref{fig:param} and from the constraint on $\gamma$ given in the preceding paragraph, and it is confirmed by a set of additional simulations (not shown) with $0.50\leq \beta_{\varphi\sigma}\leq 0.51$.  In these simulations, the region of parameter space in which a collision results in the appearance of an unstable string state (in the form of an expanding bubble) is larger for large values of $\beta_{\varphi\sigma}$, while it shrinks considerably for $\beta_{\varphi\sigma}\lesssim 0.50$.

\begin{figure}[!ht]
\includegraphics[width=8cm]{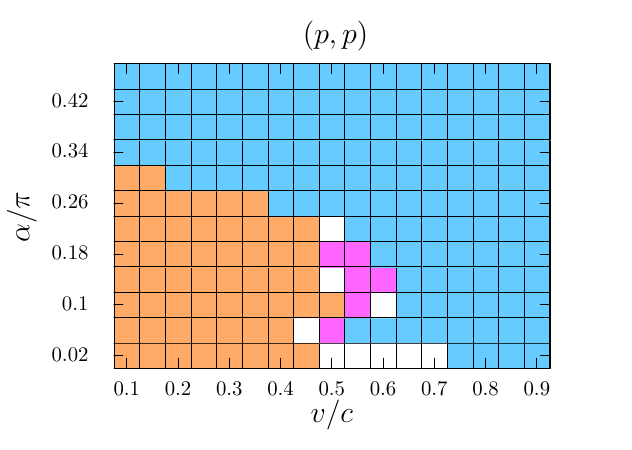}
\includegraphics[width=8cm]{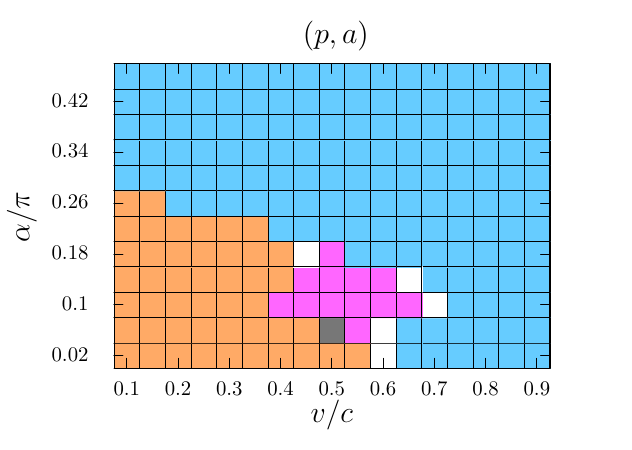}
\caption{The phase diagram of the colliding strings for Model III, with $(\beta_{\varphi},\beta_{\varphi\sigma},\beta_{\sigma},\mu_{\sigma}^2,K^2)=(1.0,0.49,1.0,0.01,0.02)$. For the correspondence between the colours in the plots and the string collision outcomes, see Table~\ref{tab:criterion}.}
\label{fig:phase3}
\end{figure}

\begin{figure}[!ht]
\includegraphics[width=8cm]{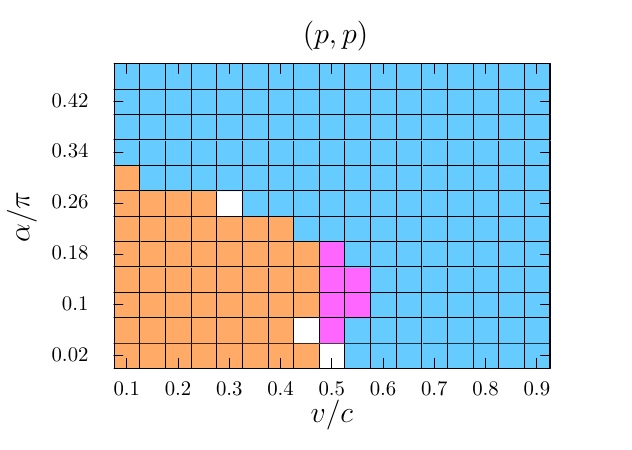}
\includegraphics[width=8cm]{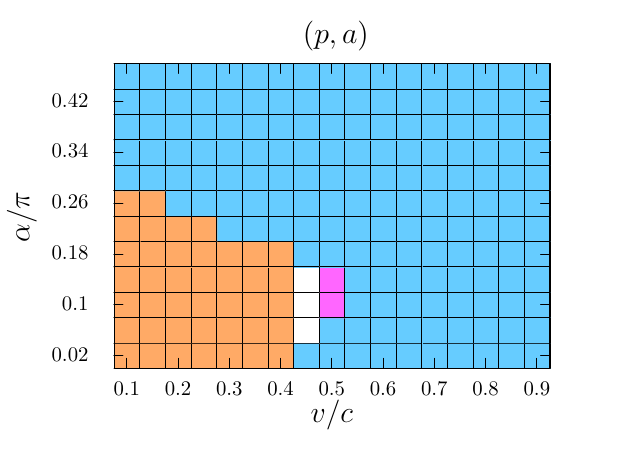}
\caption{The phase diagram of the colliding strings for Model IV, with $(\beta_{\varphi},\beta_{\varphi\sigma},\beta_{\sigma},\mu_{\sigma}^2,K^2)=(1.0,0.49,1.0,0.01,0.07)$. For the correspondence between the colours in the plots and the string collision outcomes, see Table~\ref{tab:criterion}.}
\label{fig:phase4}
\end{figure}

%%%%%%%%%%%%%%%%%%%%%%%%%
\subsubsection{Model III \& IV}
\label{subsubsec:model3}
%%%%%%%%%%%%%%%%%%%%%%%%%

In going from Model III to IV, the squared amplitude of the currents in the two colliding, $K^2$,  is increased from 0.02 to 0.07.  The corresponding phase diagrams are shown in~\cref{fig:phase3} and \cref{fig:phase4}, respectively, and are similar to the ones of Model I, see~\cref{fig:phase1}.  In particular, strings intercommute in the usual way for all $\alpha>\pi/2$. For this reason, in~\cref{fig:phase3} and~\cref{fig:phase4}, the phase plots are shown for the reduced range $0\leq \alpha \leq \pi/2$ rather than for the full range 0 to $\pi$.
The similarity of these phase plots suggests that the initial current $K$ does not play an essential role in the outcome of the collision as long as $\beta_{\varphi\sigma}<0.5$. This applies especially to the regions in which bound states form and regular intercommutation occurs. We refer the reader to the discussion in~\cref{subsubsec:model2} for the case $\beta_{\varphi\sigma}>0.5$.

The region of double intercommutation is smaller when the amplitude of the current in the colliding strings is larger. This is visible by comparing the results for Model I and IV with those of Model III.  One can conclude that larger currents hinder the occurrence of double intercommutation and instead favour regular intercommutation and/or the formation of bound states.

\begin{figure}[!ht]
\includegraphics[width=8cm]{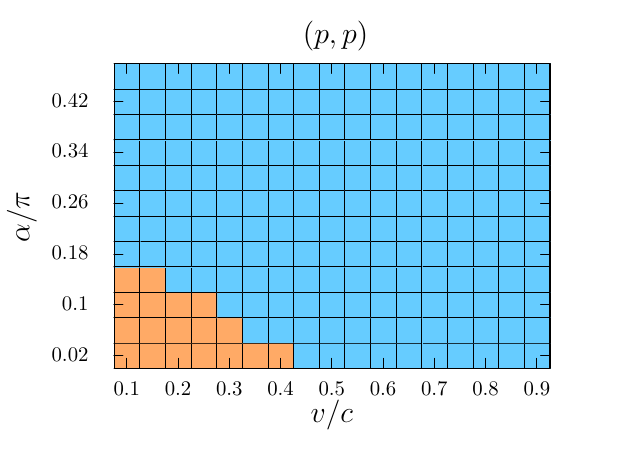}
\includegraphics[width=8cm]{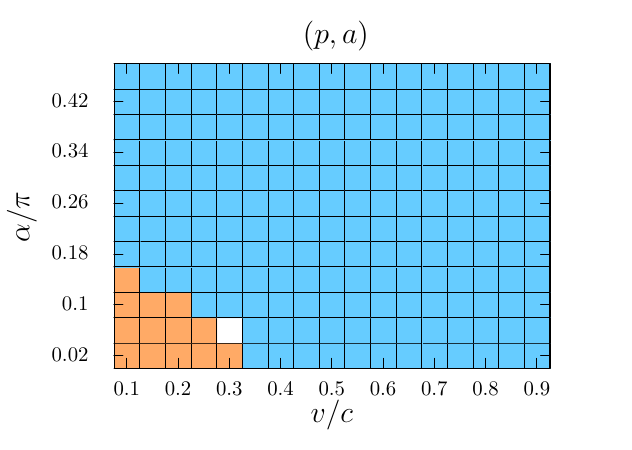}
\caption{The phase diagram of the colliding strings for Model V, with $(\beta_{\varphi},\beta_{\varphi\sigma},\beta_{\sigma},\mu_{\sigma}^2,K^2)=(1.0,0.4,1.0,0.01,0.02)$. For the correspondence between the colours in the plots and the string collision outcomes, see Table~\ref{tab:criterion}.}
\label{fig:phase5}
\end{figure}

%%%%%%%%%%%%%%%%%%%%%%%%%%
\subsubsection{Model V}
\label{subsubsec:model5}
%%%%%%%%%%%%%%%%%%%%%%%%%%%

Let us now study the influence of $\beta_{\varphi\sigma}$, the coupling
constant between $\varphi$ and $\sigma$. To do so, we compare the results of Model III in~\cref{fig:phase3} with those of Model V in~\cref{fig:phase5}. In going from Model III to Model V, $\beta_{\varphi\sigma}$ is decreased from 0.49 to 0.40, while all other parameters are kept identical.  We find that bound states form less frequently for smaller $\beta_{\varphi\sigma}$ (Model V). This result is reasonable. Indeed, if we consider the limit $\beta_{\varphi\sigma}\to 0$, the strings are similar to critical Abelian-Higgs strings, which do not form bound states. 

While in comparing Model I and III, we found that the size of the region in which bound states form is largely insensitive to $K$, here, we find, by comparing models III and V, that its size has a clear dependence on $\beta_{\varphi\sigma}$. We thus conclude that the parameter which determines whether bound states can form is the coupling $\beta_{\varphi\sigma}$.

\begin{figure}[!ht]
\includegraphics[width=8cm]{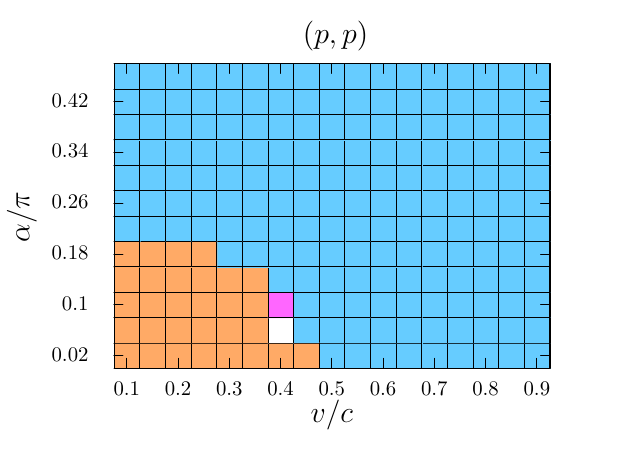}
\includegraphics[width=8cm]{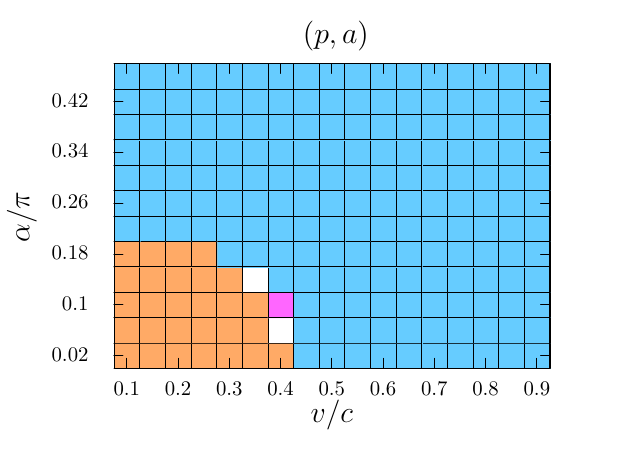}
\caption{The phase diagram for the colliding strings of Model VI with $(\beta_{\varphi},\beta_{\varphi\sigma},\beta_{\sigma},\mu_{\sigma}^2,K^2)=(1.0,0.49,1.0,0.07,0.01)$. For the correspondence between the colours in the plots and the string collision outcomes, see Table~\ref{tab:criterion}.}
\label{fig:phase6}
\end{figure}

%%%%%%%%%%%%%%%%%%%%%%%%%
\subsubsection{Model VI}
\label{subsubsec:model6}
%%%%%%%%%%%%%%%%%%%%%%%%%

Finally, let us consider a case with large $\mu_\sigma^2$, the bare mass of $\sigma$. Model IV (\cref{fig:phase4}) and VI both have $\gamma=0.08$, with the values of $K$ and $\mu_{\sigma}$ interchanged.  As explained in~\cref{subsec:static}, 
the stability of a straight and static string is determined by two parameters, $\beta_{\varphi\sigma}$ and $\gamma=K^2-\Omega^2+\mu_\sigma^2$ (recall that $\Omega=0$ in our simulations). While the individual values of $\mu_\sigma^2$ and $K^2$ do not play an essential role in the analysis of static strings, we find, by comparing the results of Model IV and Model VI, that the dynamical process does depend on $\mu_\sigma^2$ itself.

The phase diagram of Model VI is shown in~\cref{fig:phase6}. We find a significant decrease in the size of the region of kinematic parameter space in which bound states can form in Model VI, in comparison to Model IV. 
To explain this difference, we note, from \cref{eq:defV}, that the 
effective mass of $\sigma$, in the core of the string, is bounded from below by $\mu_\sigma$ ($m_\sigma$).  Correspondingly, the length scale over which two strings interact, which is roughly given by the inverse of the effective mass, is bounded from above by $\mu_\sigma^{-1}$.
%%%%%%%%%%%%%%%%%%%%%%%%%
%%%%%%%%%%%%%%%%%%%%%%%%%
\section{Conclusion}
\label{sec:conclusion}
%%%%%%%%%%%%%%%%%%%%%%%%%
%%%%%%%%%%%%%%%%%%%%%%%%%
In this work, we studied the collision process of two superconducting strings using field-theoretic simulations.  We considered a complex scalar field $\varphi(\xx,t)$ with its associated
gauge field $A_{\mu}(\xx,t)$ and another complex scalar field $\sigma(\xx,t)$. 
This model has $U_{\rm local}(1)\times U_{\rm global}(1)$ symmetry,
and the breaking of the $U_{\rm local}(1)$ symmetry leads to the formation of cosmic strings.

We first considered an Abrikosov-Nielsen-Olesen vortex solution on which the current-carrier $\sigma$ condenses. We solved the field equations numerically in order to obtain the field configuration and determine the region of parameter space in which strings are viable. The parameter space considered is the  $(\beta_{\varphi\sigma},\gamma)$ plane, where $\beta_{\varphi\sigma}=\lambda_{\varphi\sigma}/2e^2$ is the coupling constant between $\varphi$ and $\sigma$, and $\gamma=K^2-\Omega^2+\mu_\sigma^2$
is the effective mass of $\sigma$ condensed on the string.

We then considered two straight current-carrying strings in a collision process. 
From the static string solutions, we performed a Lorentz transformation in order to obtain moving strings with velocity $v$ and relative angle $\alpha$. In a three-dimensional computational domain, we then performed dynamic simulations of the collision process.
 Using a set of criteria defined in \cref{tab:criterion} we classified the final string configuration after the collision.
The results are summarised in the phase diagrams of ~\cref{fig:phase1}--\cref{fig:phase6} for the six sets of string parameters defined in \cref{tab:param}.

While most colliding strings do intercommute in the same manner as in the Abelian-Higgs model in parts of the string parameter space and collision configuration space, we also found a variety of other final states depending on the model parameters and on the kinetics of the simulations.
With a relatively small velocity, $v/c\lesssim 0.5$ and a small angle, $\alpha\lesssim 0.3\pi$, the colliding strings form a bound state; this is consistent with what can be observed in Type-I Abelian-Higgs strings. We also observe that two strings can pass through one another in a double intercommutation mechanism. This phenomenon is also observed in the collision of Type-I/II Abelian-Higgs strings at speeds close to $c$. 

Furthermore, two strings with a relatively large $\beta_{\varphi\sigma}$ form an expanding bubble after the collision, and the string configuration breaks down. In the bubble's interior, $\varphi$ and $\sigma$ lie at the extremum labelled [C], defined in~\cref{subsec:condition}. Once a small segment of string around the impact point enters the region of string instability region in the $(\beta_{\varphi\sigma},\gamma)$ plane, it expands and tends to fill the entire 3D computational domain. This phenomenon can be observed only if $\beta_{\varphi\sigma}$ is larger than its critical value.

In summary, our numerical studies demonstrate that a substantial variety of final string configurations exist in a dynamic current-carrying string collision process. This suggests that in a cosmological context, the evolution of the network of superconducting strings should be more complicated than that of Abelian-Higgs strings. It is therefore non-trivial that a superconducting string network would follow scaling laws similar to what is found in the Abelian-Higgs model~\cite{Kibble:1984hp,Martins:1995tg,Martins:1996jp,Martins:2000cs,Hiramatsu:2013tga,Hindmarsh:2017qff,Hindmarsh:2018wkp,Correia:2018gew}. 
This being said, should the superconducting string network indeed follow a scaling law, the number density of superconducting strings may differ from the one found in the Abelian-Higgs case due to the formation of bound states and double intercommutation. This could leave a distinct characteristic imprint on the cosmic microwave background and gravitational wave background. Furthermore, if standard model particles or dark matter particles condense on strings, the resulting superconducting strings could generate fast radio bursts~\cite{Vachaspati:2008su,Yu:2014gea} and could also be at the origin of gamma-ray bursts~\cite{Berezinsky:2001cp,Cheng:2010ae}.
We leave the study of superconducting string networks for a future study.

\begin{acknowledgments}
This project has been launched based on the discussion with Dani\`ele A.~Steer, and we thank her for the useful discussion that followed. This work was supported by JSPS KAKENHI Grant Numbers 16K17695, 21K03559, 23H00110 (T. H.) and 17K14304, 19H01891, 22K03627 (D. Y.).
\end{acknowledgments}

\appendix
%%%%%%%%%%%%%%%%%%%%%%%%%%%%%%%%%%%%%%%%%%%%%%%%%%%%%%%%%%%%%%%%%%%%%%%%%%%%%%%
%%%%%%%%%%%%%%%%%%%%%%%%%%%%%%%%%%%%%%%%%%%%%%%%%%%%%%%%%%%%%%%%%%%%%%%%%%%%%%%
\section{Notations}
\label{appsec:notations}
%%%%%%%%%%%%%%%%%%%%%%%%%%%%%%%%%%%%%%%%%%%%%%%%%%%%%%%%%%%%%%%%%%%%%%%%%%%%%%%
%%%%%%%%%%%%%%%%%%%%%%%%%%%%%%%%%%%%%%%%%%%%%%%%%%%%%%%%%%%%%%%%%%%%%%%%%%%%%%%

We use the same model as used in Peter's paper \cite{Peter:1992dw}, but follow
the different notations. The correspondence of the variables and
parameters between the literature and ours is as follows:
%%%%%%%%%%%%%%%%%%%%%%%%%%%%%%%%%%%%%%%%%%%%%%%%%%%%%%%%%%%%%%%%%%%%
%
\begin{gather}
 \varphi = \frac{1}{\sqrt{2}}\Phi^{\rm Peter}, \quad
 \sigma = \frac{1}{\sqrt{2}}\Sigma^{\rm Peter}, \quad
 A_\mu = -\frac{1}{\sqrt{4\pi}}B_\mu^{\rm Peter}, \\
 e = \sqrt{4\pi}q^{\rm Peter}, \quad
 \eta = \frac{1}{\sqrt{2}}\eta^{\rm Peter}, \quad
 m_\sigma = \sqrt{2}m_{\sigma}^{\rm Peter}, \\
 \lambda_\varphi = 2\lambda_{\phi}^{\rm Peter}, \quad
 \lambda_{\varphi\sigma} = 4f^{\rm Peter}, \quad
 \lambda_\sigma = 4\lambda_{\sigma}^{\rm Peter}.
\end{gather}
%
%%%%%%%%%%%%%%%%%%%%%%%%%%%%%%%%%%%%%%%%%%%%%%%%%%%%%%%%%%%%%%%%%%%%
The dimensionless parameters $\alpha_i$ ($i=1,2,3$) and $\widetilde{w}$,
$\widetilde{q}$ defined in Ref.~\cite{Peter:1992dw} can be represented
by our notations,
%%%%%%%%%%%%%%%%%%%%%%%%%%%%%%%%%%%%%%%%%%%%%%%%%%%%%%%%%%%%%%%%%%%%
%
\begin{gather}
 \alpha_1 = \frac{m_\sigma^2}{\lambda_\sigma\eta^2}
          = \frac{\mu_\sigma^2}{\beta_\sigma}, \quad
 \alpha_2 = \frac{\lambda_{\varphi\sigma}m_\sigma^2}{2\lambda_\varphi\lambda_\sigma\eta^2}
          = \frac{\beta_{\varphi\sigma}\mu_\sigma^2}{2\beta_\varphi\beta_\sigma}, \quad
 \alpha_3 = \frac{m_\sigma^4}{2\lambda_\varphi\lambda_\sigma\eta^4}
          = \frac{\mu_\sigma^4}{2\beta_\varphi\beta_\sigma}, \\
 \widetilde{q}^2 = \frac{1}{4\pi\beta_{\varphi}}, \quad
 \widetilde{w} = \frac{\mu_\sigma^2W}{2\beta_\varphi\beta_\sigma},
\end{gather}
%
%%%%%%%%%%%%%%%%%%%%%%%%%%%%%%%%%%%%%%%%%%%%%%%%%%%%%%%%%%%%%%%%%%%%
and we can rewrite
$\beta_\varphi,\beta_{\varphi\sigma},\beta_{\sigma}$ and $\mu_\sigma^2$
with $\alpha_i$ and $\widetilde{q}$.
In Ref.~\cite{Peter:1992dw}, the authors used two kinds of parameter sets,
%%%%%%%%%%%%%%%%%%%%%%%%%%%%%%%%%%%%%%%%%%%%%%%%%%%%%%%%%%%%%%%%%%%%
%
\begin{align}
 1)&\;\;\alpha_1=1.68\times 10^{-2}, \quad
 \alpha_2=5.26\times 10^{-3}, \quad
 \alpha_3=5.26\times 10^{-4}, \quad
  4\pi\widetilde{q}^2 = 0.1, \\
 2)&\;\;\alpha_1=2\times 10^{-3}, \quad
 \alpha_2=5\times 10^{-4}, \quad
 \alpha_3=2\times 10^{-6}, \quad
  4\pi\widetilde{q}^2 = 0.1.
\end{align}
%
%%%%%%%%%%%%%%%%%%%%%%%%%%%%%%%%%%%%%%%%%%%%%%%%%%%%%%%%%%%%%%%%%%%%
These respectively correspond to
%%%%%%%%%%%%%%%%%%%%%%%%%%%%%%%%%%%%%%%%%%%%%%%%%%%%%%%%%%%%%%%%%%%%
%
\begin{align}
 1)&\;\;\beta_\varphi = 10.0, \quad
 \beta_{\varphi\sigma} = 6.26, \quad
 \beta_\sigma = 37.3, \quad
 \mu_\sigma = 0.791, \\
 2)&\;\;\beta_\varphi = 10.0, \quad
 \beta_{\varphi\sigma} = 5.0, \quad
 \beta_\sigma = 10.0, \quad
 \mu_\sigma = 0.141.
\end{align}
%
%%%%%%%%%%%%%%%%%%%%%%%%%%%%%%%%%%%%%%%%%%%%%%%%%%%%%%%%%%%%%%%%%%%%
From 5 of Ref.~\cite{Peter:1992dw}, we can read 
$\widetilde{w}\sim 10^{-3}$ for the first parameter set called
'moderate' case in the main text. This corresponds to
$W\approx 1.19\left(\widetilde{w}/10^{-3}\right)$.

%%%%%%%%%%%%%%%%%%%%%%%%%%%%%%%%%%%%%%%%%%%%%%%%%%%%%%%
\section{Viable parameter region for a superconducting string}
\label{appsec:viable}
%%%%%%%%%%%%%%%%%%%%%%%%%%%%%%%%%%%%%%%%%%%%%%%%%%%%%%%

\begin{figure}[!ht]
\centering{
\includegraphics[width=8cm]{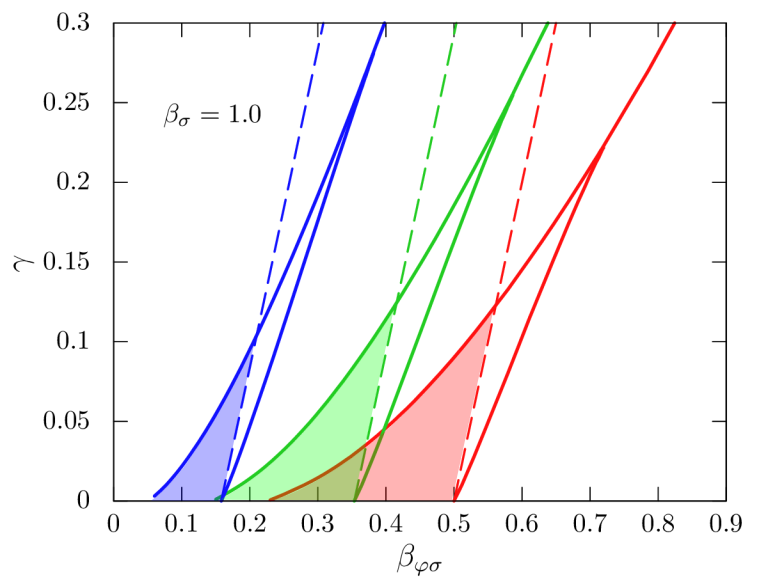}
\includegraphics[width=8cm]{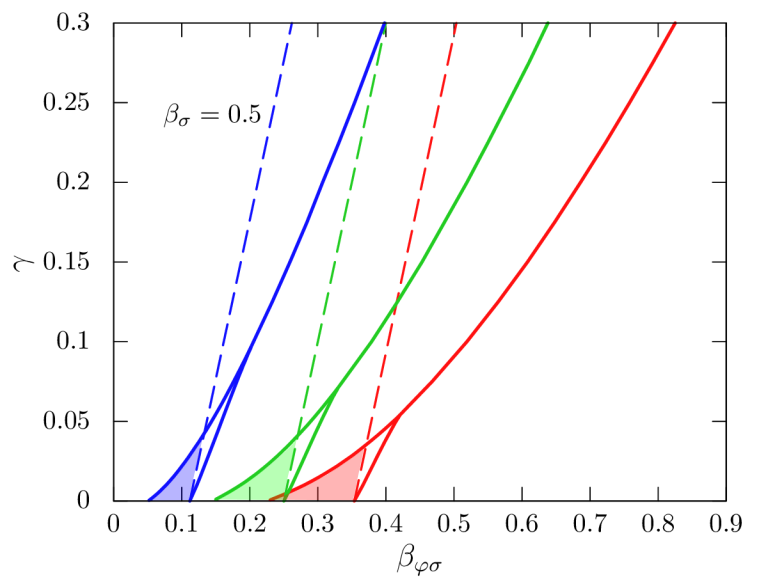}
}
\caption{Viable parameter region for a superconducting string with $\beta_{\varphi}=1.0$ (red), $0.5$ (green) and $0.1$ (blue). In the left-hand (right-hand) plot, we set $\beta_{\sigma}=1.0$ ($\beta_{\sigma}=0.5$). The shaded area indicates the region of parameter space in which superconducting strings form, i.e., the region labelled 'SC' in \cref{fig:param}. The red solid and dashed lines are the same as in \cref{fig:param}.}
\label{fig:param_all}
\end{figure}

In~\cref{subsec:static}, we explored the viable parameter region for a superconducting string in $(\beta_{\varphi\sigma},\gamma)$ space. In the left panel of~\cref{fig:param_all}, we vary $\beta_{\varphi}$ as $\beta_{\varphi}=1.0$ (red), $0.5$ (green) and $0.1$ (blue). Basically, as $\beta_{\varphi}$ decreases or $\beta_{\sigma}$ decreases, the viable region ('SC' region in~\cref{fig:param}) moves toward the left to satisfy the positivity of Eq.~(\ref{eq:cond3}). We find that the boundary of the left-hand region (``NC" region in~\cref{fig:param}) is insensitive to $\beta_{\sigma}$. We discuss this point in \cref{appsec:NCregion}.

%%%%%%%%%%%%%%%%%%%%%%%%%%%%%%%%%%%%%%%%%%%%%%%%%%%%%%%
\section{Determining the boundary of the ``NC'' region }
\label{appsec:NCregion}
%%%%%%%%%%%%%%%%%%%%%%%%%%%%%%%%%%%%%%%%%%%%%%%%%%%%%%%

We shall analytically derive the boundary of the ``NC'' region in~\cref{fig:param}. In~\cref{sec:NOV}, we considered the condition for the current-carrier to condensate on a string from the potential shape of $\widetilde{V}_{\rm eff}$. However, an actual string configuration also highly depends on the gradient energy. So we derive the condition taking into account the gradient energy by approximating the field configuration.

In the case with $\beta_{\varphi}=1$ and the winding number $n=1$, we find that
%%%%%%%%%%%%%%%%%%%%%%%%%%%%%%%%%%%%%%%%%%%%%%%%%%%%%%%%%%%%%%%%%%%%
%
\begin{align}
f(x) = \tanh\frac{x}{d}, \quad \alpha(x) = \tanh^2\frac{x}{d}, \quad g(x) = \frac{g_0}{\cosh(x/d)}.
\end{align}
%
%%%%%%%%%%%%%%%%%%%%%%%%%%%%%%%%%%%%%%%%%%%%%%%%%%%%%%%%%%%%%%%%%%%%
can approximate the numerical solutions in the right panel of~\cref{fig:conf2} with $d\approx 1.2$ and $g_0\approx 0.5$. For $\beta_{\sigma}=0.5$ as in the right panel of~\cref{fig:param_all}, $d$ changes within only $10\%$, so these ansatz work also in this case.

Plugging these ansatz with $\beta_{\varphi}=n=1$ into~\cref{eq:red_Lag}, the Lagrangian becomes
%%%%%%%%%%%%%%%%%%%%%%%%%%%%%%%%%%%%%%%%%%%%%%%%%%%%%%%%%%%%%%%%%%%%
%
\begin{align}
-\frac{\mathcal{L}}{e^2\eta^4} &= \frac{x}{2d^2}\left(2+d^2+d^2\beta_{\sigma}g_0^4+g_0^2(-1+d^2(-4 \beta_{\varphi\sigma}+\gamma))+g_0^2(1+d^2\gamma)\cosh\frac{2x}{d}\right){\rm sech}^4\frac{x}{d}
\notag \\ &\quad
+\left(1+\frac{2}{d^2}\right)\frac{1}{x} {\rm sech}^4\frac{x}{d}\tanh^2\frac{x}{d}.
\end{align}
%
%%%%%%%%%%%%%%%%%%%%%%%%%%%%%%%%%%%%%%%%%%%%%%%%%%%%%%%%%%%%%%%%%%%%
The tension of a superconducting string, the energy per the unit length, is given by integrating the Lagrangian over $x=[0,\infty]$. The integral of the last term cannot be expressed in terms of elementary functions, which gives only a constant contribution to $\mu$. Then we obtain the tension of a superconducting string,
%%%%%%%%%%%%%%%%%%%%%%%%%%%%%%%%%%%%%%%%%%%%%%%%%%%%%%%%%%%%%%%%%%%%
%
\begin{align}
\mu = 2\pi \int\!\mathcal{L} xdx
 = 2\pi\left( c + a_2g_0^2 + a_4 g_0^4 \right),
\end{align}
%
%%%%%%%%%%%%%%%%%%%%%%%%%%%%%%%%%%%%%%%%%%%%%%%%%%%%%%%%%%%%%%%%%%%%
where $c$ is constant and
%%%%%%%%%%%%%%%%%%%%%%%%%%%%%%%%%%%%%%%%%%%%%%%%%%%%%%%%%%%%%%%%%%%%
%
\begin{align}
a_2 &= \frac{\pi}{3}\left( 1+2\ln 2 - 2d^2 \left((4\ln 2-1)\beta_{\varphi\sigma}-3\ln 2\gamma\right)\right), \\
a_4 &= \frac{\pi}{6}\left(4\ln 2-1\right)\beta_{\sigma}d^2>0.
\end{align}
%
%%%%%%%%%%%%%%%%%%%%%%%%%%%%%%%%%%%%%%%%%%%%%%%%%%%%%%%%%%%%%%%%%%%%
Requiring the minimum of the tension to exist, we obtain the condition, $a_2<0$, which yields
%%%%%%%%%%%%%%%%%%%%%%%%%%%%%%%%%%%%%%%%%%%%%%%%%%%%%%%%%%%%%%%%%%%%
%
\begin{align}
 2d^2\beta_{\varphi\sigma}(4\ln 2-1)  - (2\ln 2+1) > 6d^2\gamma \ln 2.
 \label{eq:NCNScond}
\end{align}
%
%%%%%%%%%%%%%%%%%%%%%%%%%%%%%%%%%%%%%%%%%%%%%%%%%%%%%%%%%%%%%%%%%%%%
This equation with $d=1.2$ can well reproduce the boundary of the 'NC' region in~\cref{fig:param}.
Too small $\beta_{\varphi\sigma}$ cannot satisfy this condition, preventing the condensation of the current-carrier on the string, namely, $g_0=0$. Clearly, the simplest analysis taking into account only the potential shape discussed in~\cref{sec:NOV} is not enough to explain the location of the boundary. The condensation is an outcome of the balance of the potential energy and the gradient energy.

In addition, this condition is independent to $\beta_{\sigma}$. The effect appears only through the weak dependence of $d$. This result is consistent with the numerical result in~\cref{fig:param_all}, where the location of the boundary is insensitive to $\beta_{\sigma}$ .

This analysis can be applied to the case with $\beta_{\varphi\sigma}\lesssim 1$.
From the numerical analysis with $\beta_{\varphi\sigma}\gtrsim 1$, we find that the boundary deviates from the condition \cref{eq:NCNScond}, implying that the details of the field configuration become important for the condensation.

%%%%%%%%%%%%%%%%%%%%%%%%%%%%%%%%%%%%%%%%%%%%%%%%%%%%%%%
\section{Moving strings and field superposition}
\label{appsec:moving}
%%%%%%%%%%%%%%%%%%%%%%%%%%%%%%%%%%%%%%%%%%%%%%%%%%%%%%%

We construct a moving string by taking the Lorentz-boost of the 1-vortex solution given in~\cref{subsec:static}. We consider a string rotated by $\theta_x$ around the $x$-axis, where $\theta_x=0$ corresponds to a string along the $z$-axis. The string moves with velocity $v$ along the $x$-axis. Suppose the scalar fields and the gauge field in the static frame of the string are given as $\{\varphi(\xx),A_\mu(\xx),\sigma(\xx)\}$. Those in the observer frame, where the string is rotated around the $x$-axis and Lorentz-boosted, $\{\varphi'(\xx'),A'_\mu(\xx'),\sigma'(\xx')\}$, are given by
%%%%%%%%%%%%%%%%%%%%%%%%%%%%%%%%%%%%%%%%%%%%%%%%%%%%%%%%%%%%%%%%%%%%
%
\begin{equation}
\varphi'(\xx') = \varphi(\xx), \quad A'_\mu(\xx') = 
G^\nu{}_\mu(-\beta;-\theta_x)A_\nu(\xx),
\quad \sigma'(\xx') = \sigma(\xx),
\end{equation}
%
%%%%%%%%%%%%%%%%%%%%%%%%%%%%%%%%%%%%%%%%%%%%%%%%%%%%%%%%%%%%%%%%%%%%
with $\beta=v/c$ and
%%%%%%%%%%%%%%%%%%%%%%%%%%%%%%%%%%%%%%%%%%%%%%%%%%%%%%%%%%%%%%%%%%%%
%
\begin{equation}
x^\mu{}' = G^\mu{}_\nu(\beta;\theta_x) x^\nu{}.
\end{equation}
%
%%%%%%%%%%%%%%%%%%%%%%%%%%%%%%%%%%%%%%%%%%%%%%%%%%%%%%%%%%%%%%%%%%%%
with the matrix $G^\mu{}_\nu$ defined as
%%%%%%%%%%%%%%%%%%%%%%%%%%%%%%%%%%%%%%%%%%%%%%%%%%%%%%%%%%%%%%%%%%%%
%
\begin{equation}
G(\beta;\phi_x) \equiv 
   \Lambda_x(\beta)R_x(\theta_x),
\end{equation}
%
%%%%%%%%%%%%%%%%%%%%%%%%%%%%%%%%%%%%%%%%%%%%%%%%%%%%%%%%%%%%%%%%%%%%
where
%%%%%%%%%%%%%%%%%%%%%%%%%%%%%%%%%%%%%%%%%%%%%%%%%%%%%%%%%%%%%%%%%%%%
%
\begin{equation}
 \Lambda_x(\beta) = 
 \begin{pmatrix}
  1/\sqrt{1-\beta^2} & \beta/\sqrt{1-\beta^2} & 0 & 0 \\
  \beta/\sqrt{1-\beta^2} & 1/\sqrt{1-\beta^2} & 0 & 0 \\
  0 & 0 & 1 & 0 \\
  0 & 0 & 0 & 1 
 \end{pmatrix},\quad
 R_x(\theta_x) = 
 \begin{pmatrix}
  1 & 0 & 0 & 0 \\
  0 & 1 &  0          & 0 \\
  0 & 0 &  \cos\theta_x & -\sin\theta_x \\
  0 & 0 &  \sin\theta_x & \cos\theta_x
 \end{pmatrix}.
\end{equation}
%
%%%%%%%%%%%%%%%%%%%%%%%%%%%%%%%%%%%%%%%%%%%%%%%%%%%%%%%%%%%%%%%%%%%%

The field theory's computational domain is defined in the observer frame, $\xx'$. Let us now the prime, and from hereon denote $\xx$ as the observer frame. At time $t=0$, the 2-vortex solution in the observer frame is given as the superposition of each 1-vortex solution \cite{VilenkinShellard},
%%%%%%%%%%%%%%%%%%%%%%%%%%%%%%%%%%%%%%%%%%%%%%%%%%%%%%%%%%%%%%%%%%%%
%
\begin{equation}
 \varphi(\xx) = \frac{1}{\eta}\varphi^{(1)}(\xx-\xx^{(1)}_0)\varphi^{(2)}(\xx-\xx^{(2)}_0), \quad
 A_\mu(\xx) = A_\mu^{(1)}(\xx-\xx^{(1)}_0)+A_\mu^{(2)}(\xx-\xx^{(2)}_0), \quad
 \sigma(\xx) = \sigma^{(1)}(\xx-\xx^{(1)}_0)+\sigma^{(2)}(\xx-\xx^{(2)}_0),
 \label{eq:multi}
\end{equation}
%
%%%%%%%%%%%%%%%%%%%%%%%%%%%%%%%%%%%%%%%%%%%%%%%%%%%%%%%%%%%%%%%%%%%%
where the superscript $(i)$ labels each moving vortex. The superposition of fields with different Lorentz transformations is justified by the fact that the Lorentz gauge condition is, of course, Lorentz invariant. At the initial time, vortex (1) and vortex (2) are sufficiently distant from one another, $|\xx^{(1)}_0-\xx^{(2)}_0|=20/(e\eta)$, so that their interactions can be neglected.

%%%%%%%%%%%%%%%%%%%%%%%%%%%%%%%%%%%%%%%%%%%%%%%%%%%%%%%
\section{Optimal grid size}
\label{appsec:gridsize}
%%%%%%%%%%%%%%%%%%%%%%%%%%%%%%%%%%%%%%%%%%%%%%%%%%%%%%%

We briefly explain how we determine the size of the computational domain for the colliding strings.
The box size depends on the collision velocity $v$, the collision angle $\alpha$ and how long we want to follow the time-evolution of the strings after the collision.

If we require that the strings with a small angle $\alpha$ are sufficiently far apart from each other at the top and bottom of the computational domain, one can imagine that the box is elongated in the $z$-direction. If the velocity $v$ is large, the strings also need sufficient volume in the $x$-direction, as the strings will travel a long distance in the $x$-direction. Also, after the collision, the aftermath travels along the strings towards outside the collision point. The simulation must be completed before this aftermath reaches the edge of the computational domain. Therefore, fixing the simulation time determines the size of the entire computational domain.

In~\cref{fig:domain}, we show the 3D computational domain.
The strings are placed on the planes parallel to the $\Sigma_{x}$ plane at $x=\pm d$.
The length of the box along the $x$-axis, $L_x$, is given as
%%%%%%%%%%%%%%%%%%%%%%%%%%%%%%%%%%%%%%%%%%%%%%%%%%%%%%%%%%%%%%%%%%%%
%
\begin{align}
 L_x = {\rm max}(2v t_{\rm end}-2d,~2d) + 2m_x,
\end{align}
%
%%%%%%%%%%%%%%%%%%%%%%%%%%%%%%%%%%%%%%%%%%%%%%%%%%%%%%%%%%%%%%%%%%%%
where $m_x$ is the constant margin and $t_{\rm end}$ is
the simulation time.
Here we assume that the aftermath of the impact propagates on the string at the speed of light.
Then the simulation time $t_{\rm end}$ is related to the velocity $v$ and the propagation distance of the aftermath $\ell_{s0}$ as
%%%%%%%%%%%%%%%%%%%%%%%%%%%%%%%%%%%%%%%%%%%%%%%%%%%%%%%%%%%%%%%%%%%%
%
\begin{align}
 t_{\rm end} = \frac{d}{v} + \frac{\ell_{s0}}{c}.
\end{align}
%
%%%%%%%%%%%%%%%%%%%%%%%%%%%%%%%%%%%%%%%%%%%%%%%%%%%%%%%%%%%%%%%%%%%%

If the length of the box along the $z$-axis is $L_z$ and the minimum separation of the strings at $z=\pm L_{z}/2$ is $s_{\rm min}$, we obtain the relation,
%%%%%%%%%%%%%%%%%%%%%%%%%%%%%%%%%%%%%%%%%%%%%%%%%%%%%%%%%%%%%%%%%%%%
%
\begin{align}
 L_z \tan\frac{\alpha}{2} > s_{\rm min}.
  \label{eq:Lz_smin}
\end{align}
%
%%%%%%%%%%%%%%%%%%%%%%%%%%%%%%%%%%%%%%%%%%%%%%%%%%%%%%%%%%%%%%%%%%%%
If $s_{\rm min}$ is too small, the collision between strings can suddenly occur across the entire computational domain. The aftermath of the collision propagates along the strings. Its distance along the $z$-axis should be less than $L_{z}$,
%%%%%%%%%%%%%%%%%%%%%%%%%%%%%%%%%%%%%%%%%%%%%%%%%%%%%%%%%%%%%%%%%%%%
%
\begin{align}
 L_{z} > 2\ell_{s0}\cos\frac{\alpha}{2}.
   \label{eq:Lz_ell}
\end{align}
%
%%%%%%%%%%%%%%%%%%%%%%%%%%%%%%%%%%%%%%%%%%%%%%%%%%%%%%%%%%%%%%%%%%%%
Combining \cref{eq:Lz_smin} and \cref{eq:Lz_ell}, we determine $L_{z}$ from

%%%%%%%%%%%%%%%%%%%%%%%%%%%%%%%%%%%%%%%%%%%%%%%%%%%%%%%%%%%%%%%%%%%%
%
\begin{align}
 L_z &= \max\left\{ \frac{s_{\rm min}}{\tan(\alpha/2)},~2\ell_{s0}\cos\frac{\alpha}{2}\right\}.
\end{align}
%
%%%%%%%%%%%%%%%%%%%%%%%%%%%%%%%%%%%%%%%%%%%%%%%%%%%%%%%%%%%%%%%%%%%%
Finally, the length of the box along the $y$-axis is given by
%%%%%%%%%%%%%%%%%%%%%%%%%%%%%%%%%%%%%%%%%%%%%%%%%%%%%%%%%%%%%%%%%%%%
%
\begin{align}
L_y &= L_z \tan\frac{\alpha}{2}+2m_y,
\end{align}
%
%%%%%%%%%%%%%%%%%%%%%%%%%%%%%%%%%%%%%%%%%%%%%%%%%%%%%%%%%%%%%%%%%%%%
where $m_y$ is the constant margin.

Throughout this paper, we set 
$m_x=15/(e\eta), m_y=15/(e\eta), s_{\rm min}=10/(e\eta)$ 
and $\ell_{s0}=100/(e\eta)$. For some cases whose final configuration
cannot be clearly judged, we set $\ell_{s0}=300/(e\eta)$ to perform
simulations for a longer time. From this procedure, the typical number of
grid is $202\leq N_x\leq 482, 162\leq N_y\leq 914, 302\leq N_z\leq 1274$
with the spatial interval $\Delta x=0.25/(e\eta)$.

\bibliographystyle{apsrev4-1}
\bibliography{master}

\end{document}